\documentclass[aps,preprint,eqsecnum,groupedaddress,nofootinbib]{revtex4}
\usepackage{mathrsfs}
\usepackage{amsmath}
\usepackage{amssymb}

\makeatletter

\usepackage{graphicx}

\newcommand{\be}{\begin{equation}}
\newcommand{\ee}{\end{equation}}
\newcommand{\bea}{\begin{eqnarray}}
\newcommand{\eea}{\end{eqnarray}}

\makeatother

\begin{document}
\leftline{KCL-PH-TH-2016-40}

\title{Magnetic Monopoles from Global Monopoles in the presence
of a Kalb-Ramond Field}

\author{Nick E. Mavromatos}

\affiliation{Theoretical Particle Physics and Cosmology Group, Department of Physics, King's College London, Strand, London
WC2R 2LS, UK}

\author{{{}Sarben Sarkar}}

\affiliation{Theoretical Particle Physics and Cosmology Group, Department of Physics, King's College London, Strand, London
WC2R 2LS, UK}

\begin{abstract}
 A classical solution for electromagnetic monopoles
induced by gravitational (global) monopoles in the presence of a (four-dimensional) Kalb-Ramond axion field  is
found. The magnetic charge of such a solution is induced by a non-zero Kalb-Ramond field strength, prevalent in string theory. 
Bounds from the current run of the LHC experiments are used
to constrain the parameters of the model. Because the production mechanism depends on the details of the model and its ultraviolet completion, such bounds, presently, are  only indicative.
\end{abstract}

\maketitle

\section{Introduction \label{s1} }

The existence of magnetic monopoles has been
a key question for nearly a century. The current 
experiments at the LHC, including MoEDAL (which is designed specifically to search for magnetic monopoles and other highly ionising particles), have started to provide interesting new bounds for the mass~\cite{monodata} of such messengers of new physics . Consequently it is timely to examine new ways that monopoles may manifest themselves and the possibility for their detection
by MoEDAL and other LHC experiments. We show the novel and surprising possibility that, in four dimensional spacetime, gravitation in the presence of Maxwell and Kalb-Ramond axion fields  (the latter being the dual of the field strength of a spin-one antisymmetric tensor field in the massless gravitational multiplet of  string theories~~\cite{Hammond:2002rm,string})
can lead to a magnetic monopole with strength determined by the Kalb-Ramond charge.

In the 1873 work of Maxwell, magnetic monopoles did not appear
in the magnetic Gauss' law since Nature had electric monopoles but
no magnetic monopoles. This asymmetry has puzzled physicists as far back as P. Curie. Indeed, the formulation of electromagnetism in terms
of a non-singular 4-vector potential $A_{\mu}$ requires the magnetic
induction $\vec{B}$ to be divergence free and no monopole
is allowed. Dirac~\cite{diracmono} showed that a monopole is possible with a \emph{singular}
gauge potential. He considered the magnetic field from a solenoid
in the limit of an arbitrarily thin semi-infinite solenoid. In this
limit the vanishingly small solenoid became the Dirac string. Such
a string cannot be detected through the Aharonov-Bohm effect once
the magnetic charge $g$ and the electric charge $g$ satisfy (in
natural units) $eg=\frac{n}{2}$ where $n$ is a positive integer.
The end of the solenoid becomes the monopole . The energy of the monopole
is not finite. 

A major paradigm shift in the theory of monopoles was initiated
independently by 't Hooft and Polyakov~\cite{hpmono}. They considered a model 
due to Georgi and Glashow~\cite{gg} 
which is a field theory with spontaneously broken gauge symmetry.
The non-Abelian gauge group of the Georgi-Glashow model is $SU\left(2\right)$. This gauge symmetry is spontaneously broken
down to the $U\left(1\right)$ gauge group of electromagnetism by
using a scalar field in the adjoint representation. Monopole solutions
with finite energy and quantised magnetic charges were found.

Quite recently the paradigm of large extra dimensions~\cite{largedim} has
lowered the Planck scale of gravitational physics to the order of
TeV and so in principle gravitational effects may become observable
at the LHC. In particular micro-black holes can be produced
and decay rapidly. The magnetic monopoles discussed by 'tHooft and Polyakov arise in gauge theories in the absence of gravity. Global gravitational (non-magnetic)
monopoles have been found as classical solutions of a coupled system
of gravity and a self-interacting scalar field in the adjoint representation
of a global O(3) group, but in the absence of a gauge field~\cite{vilenkin}. The global monopole is a solution for the gravitational field 
similar to a Schwarzchild black hole  with an asymptotic space-time which is Minkowski but with a deficit angle. 
A necessary
condition for a gravitational monopole to behave also as a \emph{magnetic} monopole configuration
is to couple covariantly a local $U\left(1\right)$ gauge field strength to gravity.
Calculations show
that this is not a sufficient condition to determine whether a \emph{magnetic} monopole is induced. Since in our model the scalar field is not related to electro-weak symmetry breaking, on phenomenological grounds the symmetry breaking parameter can be chosen to allow a monopole with a mass accessible to the LHC, without the need to invoke large extra dimensions.  

In some (closed) string theories~\cite{string}, a $2-$form gauge field, the spin-1 Kalb-Ramond (KR)
gauge field, appears in the massless spectrum. It is well known that, for the bosonic \emph{gravitational} part of low-energy string effective actions, the
Kalb-Ramond field strength, which the string effective actions depend upon, on account of the Kalb-Ramond gauge invariance, can be thought of as providing a source of torsion~\cite{Hammond:2002rm}.  Recently~\cite{fermi2}, in string-inspired effective theories, we have considered some cosmological implications of a dual formulation of a time-dependent four-dimensional Kalb-Ramond  field, in connection with the generation of matter-antimatter asymmetry in the Universe.  
In four space-time dimensions, the dual of the Kalb-Ramond field strength is a pseudoscalar axion-like field.
This formulation will be used here. 

We will investigate the role of \emph{static} configurations of the (dual of the) Kalb-Ramond field strength in inducing monopole solutions with a non-trivial magnetic charge. 
Our effective field theory contains the gravitational metric tensor,
a triplet of scalar fields in the adjoint representation
of the $O(3)$ group (necessary for the the spontaneous breaking of the $O(3)$ symmetry),
a local $U\left(1\right)$ $2-$form, the electromagnetic field strength,
 and a (static) $3-$form, the Kalb-Ramond  field strength. It is also \emph{necessary} to introduce into the model an additional $O(3)$-singlet scalar field, which is stabilised to a constant value. In the context of  string theory, this is the dilaton (spin zero part of the gravitational massless string multiplet), and in principle its stabilisation could be guaranteed by an appropriate (string-loop induced) dilaton potential.  However, one may imagine phenomenological scenarios independent of string theory, in which this extra scalar is ultraheavy,  is stabilised by its own potential, and is coupled only gravitationally to the other scalar and gauge fields of the model. 
For this model there is a solution whereby  the magnetic charge of the monopole is determined by the strength of the Kalb-Ramond field~\footnote{It has been shown in \cite{torsionglobal}, that the structure of the global (non-magnetic) monopole~\cite{vilenkin} remains intact in the presence of the Kalb-Ramond field. Our model differs by the inclusion of a $U(1)$-gauge field, antisymmetric tensor degrees of freedom and a singlet scalar field; the non-trivial Kalb-Ramond field strength determines the magnetic charge of the monopole. In  simple string theory sigma models with just lowest-order graviton and antisymmetric tensor fields, the Kalb-Ramond field strength can be absorbed as torsion inside a generalised scalar curvature. However in the presence of other fields this is not the case; so, in four space-time dimensions, we consider the Kalb-Ramond field to be a massless axion-like field, and  the gravitational part of our Lagrangian is kept torsion free. }. As we shall show below, within the context of string theory, it is the dilaton equation of motion that provides the link between the electromagnetic and the Kalb-Ramond field strengths. This link leads to the  connection between  the magnetic and the ``Kalb-Ramond torsion'' charges.

Since our treatment is inspired by both the 't Hooft-Polyakov
(HP) monopole solution~\cite{hpmono}  and the (self-gravitating) non-magnetic global monopole of \cite{vilenkin}, we will briefly review the main features of these
solutions in Sec.~\ref{sec:2}. This will be followed in Sec.~\ref{sec:3} by an introduction
of the Lagrangian for our model, a derivation of the coupled classical
equations of the model, and an asymptotic
analysis of the equations of the model for small and large (radial) distances from the monopole centre. 
We shall demonstrate analytically the existence of magnetic monopole solutions in these two regimes;
we estimate the monopole mass, which agrees in order of magnitude with the non-magnetic global monopole of \cite{vilenkin}. The concluding Section \ref{sec:5} discusses the phenomenology of the magnetic monopole solution, and makes some conjectural remarks 
on the possibility of its production and detection at the LHC.

\section{The 't Hooft-Polyakov and Global monopole solutions \label{sec:2}}

 We will review the basic features of the HP~\cite{hpmono} and global monopole~\cite{vilenkin} solutions, which are relevant for our model. 
We commence with the original HP monopole within the context of an $SU(2)$ spontaneously broken gauge theory with  adjoint ``Higgs'' triplet fields. Such solutions can be generalised to Grand Unified Theory (GUT) groups, such as $SU(5)$, leading to realistic particle phenomenology, and providing GUT monopoles with masses near the GUT scale ($\sim 10^{14}-10^{15}$~GeV). Such cosmic monopoles are expected to have been diluted by inflation. 

\subsection{The `t Hooft-Polyakov $SU(2)$ Monopole}
 
The fields in the HP $SU(2)$-gauge-theory model~\cite{hpmono}  are a scalar field $\phi^{a}\left(t,\vec{x}\right)$
and gauge field $A_{\mu}^{a}\left(t,\vec{x}\right)$ where
$a\left(=1,2,3\right)$ is a $SU\left(2\right)$ index. The Lagrangian
density \emph{{}$\mathscr{L\left(\mathit{\mathrm{\mathit{t,\vec{x}}}}\right)}$}
is 
\begin{equation}
\mathscr{L}\left(t,\vec{x}\right)=-\frac{1}{4}F_{\mu\nu}^{a}F^{a\mu\nu}+\frac{1}{2}\left(D_{\mu}\phi^{a}\right)\left(D^{\mu}\phi^{a}\right)-\frac{1}{4}\lambda\left(\phi^{a}\phi^{a}-\eta^{2}\right)^{2}.\label{eq:1}
\end{equation}
The field tensor $F_{\mu\nu}^{a}$ is
\begin{equation}
F_{\mu\nu}^{a}=\partial_{\mu}A_{\nu}^{a}-\partial_{\nu}A_{\mu}^{a}+g\epsilon^{abc}A_{\mu}^{b}A_{\nu}^{c}\label{eq:2}
\end{equation}
where $\epsilon^{abc}$ is the anti-symmetric Levi-Civita symbol;
$D_{\mu}\phi^{a}$ is defined by 
\begin{equation}
D_{\mu}\phi^{a}=\partial_{\mu}\phi^{a}+g\epsilon^{abc}A_{\mu}^{b}\phi^{c}\label{eq:3}
\end{equation}
is the covariant derivative. The model parameters are $g,\lambda>0$
and $\eta$. (The covariant derivative of $F_{\mu\nu}^{a}$ is defined
in an analogous fashion.) The equations of motion that follow from $\mathscr{L}$
are 
\begin{equation}
D_{\mu}F^{a\mu\nu}=g\epsilon^{abc}\left(D^{\nu}\phi^{b}\right)\phi^{c}\label{eq:4}
\end{equation}
and 
\begin{equation}
D_{\mu}D^{\mu}\phi^{a}=-\lambda\left(\phi^{b}\phi^{b}\right)\phi^{a}+\lambda\eta^{2}\phi^{a}.\label{eq:5}
\end{equation}
The ansatz used for a static solution of (\ref{eq:4}) and (\ref{eq:5})
(in the gauge $A_{0}^{a}\left(\vec{x}\right)=0)$ is 
\begin{equation}
\phi^{a}\left(\vec{x}\right)=\delta_{ia}\left(\frac{x^{i}}{r}\right)\frac{F\left(r\right)}{r}\label{eq:6}
\end{equation}
and 
\begin{equation}
A_{i}^{a}\left(\vec{x}\right)=\epsilon_{aij}\left(\frac{x^{j}}{r}\right)W\left(r\right)\label{eq:7}
\end{equation}
where $a,i,j=1,2,3$ and $r=\left|\vec{x}\right|$. Furthermore
the boundary conditions adopted are 
\begin{equation}
F\left(r\right)\rightarrow\eta\;\mathrm{and}\; W\left(r\right)\rightarrow1/gr\label{eq:8}
\end{equation}
as $r\rightarrow\infty.$ 't Hooft and Polyakov found that 
\begin{equation}
grW\left(r\right)=1-\frac{rg\eta}{\sinh\left(g\eta r\right)}\;\mathrm{and}\; grF\left(r\right)\sim \frac{rg\eta}{\tanh\left(g\eta r\right)}-1.\label{eq:9}
\end{equation}
The electromagnetic field tensor $f_{\mu\nu}$ is defined to be 
\begin{equation}
f_{\mu\nu}=\hat{\phi}^{a}F_{\mu\nu}^{a}-\frac{1}{g}\epsilon^{abc}\hat{\phi}^{a}D_{\mu}\hat{\phi}^{b}D_{\nu}\hat{\phi}^{c}\label{eq:10}
\end{equation}
where $\hat{\phi}^{a}=\phi^{a}/\left|\vec{\phi}\right|$
and $\left|\vec{\phi}\right|=\left(\sum_{a=1}^{3}\phi^{a}\phi^{a}\right)^{1/2}.$
The magnetic induction, determined by $B_{k}=\frac{1}{2}\epsilon_{kij}f_{ij},$
has an asymptotic behaviour 
\begin{equation}
\vec{B}\left(\vec{x}\right)\rightarrow\vec{x}/gr^{3}\label{eq:11}
\end{equation}
as $r\rightarrow\infty$ which corresponds to a magnetic monopole
of strength $1/g$. Moreover 
\begin{equation}
\frac{1}{2}\epsilon_{\mu\nu\rho\sigma}\partial^{\nu}f^{\rho\sigma}=\frac{1}{2g}\epsilon_{\mu\nu\rho\sigma}\epsilon_{abc}\partial^{\nu}\hat{\phi}^{a}\partial^{\rho} \, \hat{\phi}^{b} \partial^{\sigma} \, \hat{\phi}^{c}\equiv \frac{k_{\mu}}{g}\label{eq:12}
\end{equation}
where $k_{\mu}$ is a topological current. The topological charge $Q=\int d^{3}x\, k_{0}$
is quantised to be an integer $n$ and the monopole charge is $n/g$.
The HP monopole has $n=1$. It should be noted that $f_{_{\mu\nu}}$ does not satisfy
the Bianchi identity.

\subsection{The (self-gravitating) $O(3)$ Global Monopole Solution}

The scalar fields in the $O(3)$ global monopole solution of Barriola and Vilenkin~\cite{vilenkin} (BV) also form a triplet  $\chi^a$, $a=1,2,3$ which parametrise the spontaneous breaking of a global  $O(3)$ symmetry down to a global $U(1)$, by means of an appropriate potential, in which the scalar field triplet acquires a non-trivial vacuum expectation value $\eta$. 
Moreover the model was embedded into Einstein gravity. The Lagrangian of the model is given by 
\begin{equation}
L=\left(-g\right)^{1/2}\left\{ \frac{1}{2}\partial_{\mu}\chi^{a}\partial^{\mu}\chi^{a}-\frac{\lambda}{4}\left(\chi^{a}\chi^{a}-\eta^{2}\right)^{2} -R\right\} \label{eq:gm}
\end{equation}
where $g_{\mu\nu}$ is the (four space-time dimensional) metric tensor,  $g=\det\left(g_{\mu\nu}\right)$ its determinant  and $R$ is the Ricci scalar for
$g_{\mu\nu}$~\footnote{Our conventions and definitions throughout this work are: $(+,-,-,-)$ for the signature of the metric, the Riemann tensor is defined as 
$R^\lambda_{\,\,\,\,\mu \nu \sigma} = \partial_\nu \, \Gamma^\lambda_{\,\,\mu\sigma} + \Gamma^\rho_{\,\, \mu\sigma} \, \Gamma^\lambda_{\,\, \rho\nu} - (\nu \leftrightarrow \sigma)$, 
and the Ricci tensor and scalar are given by  $R_{\nu\alpha} = R^\lambda_{\,\,\,\,\nu \lambda \alpha}$ and $R= g^{\mu\nu}\, R_{\mu\nu}$ respectively.}.

As a result of the Goldstone theorem, such monopoles have massless Goldstone fields associated with them, which have energy densities that scale like $1/r^2$ with the radial distance from the monopole core. This results in a linear divergence of the monopole total energy density (that is mass), which is a characteristic feature of such solutions, in a way similar to the linearly divergent energy of a cosmic string. In the original work of \cite{vilenkin} only estimates of the total monopole mass have been given by considering the solution in the exterior of the monopole core, whose size in flat space time has been estimated to be of order $\delta \sim \lambda^{-1/2} \, \eta^{-1}$, leading to a heuristic mass estimate of order 
$M_{\rm core} \sim  \delta^3 \, \lambda \, \eta^4 = \lambda^{-1} \eta $.  The presence of the monopole curves the space-time exterior, and these estimates have to be rethought. However, the main argument of \cite{vilenkin} was that gravitational effects are weak for $\eta \ll M_{\rm P}$, the Planck mass; this is certainly the case  of interest for $\eta$ of order of a few TeV, the case of relevance to new physics searches at LHC. 
In this sense, BV argued that the flat space-time estimates for the core mass might still be valid, as an order of magnitude estimate. Outside the monopole core, BV used approximate asymptotic analysis of the Einstein equations , 
\begin{equation}
R_{\mu\nu} - \frac{1}{2} g_{\mu\nu} R = 8\pi G_{N} \, T^\chi_{\mu\nu}
\end{equation}
where $T_{\mu\nu}^\chi$ is the matter stress tensor derived from the Lagrangian (\ref{eq:gm}), and the equations of motion for the scalar fields $\chi^a$, $a=1,2,3$ . The scalar field configuration for a global monopole is~\cite{vilenkin}
\begin{equation}\label{eq:gmf}
\chi^a = \eta \, f(r) \, \frac{x^a}{r}~, a=1,2,3 
\end{equation}
where $x^a$ are spatial Cartesian coordinates, $r = \sqrt{x^a x^a }$ is the radial distance, and $f(r) \to 1 $ for  $r \gg \delta $. So 
at such large distances, the amplitude squared of the scalar field triplet approaches the square of the vacuum expectation value $\eta$, $\chi^a \chi^a \to \eta^2$. 
The reader should note the similarity between the expression (\ref{eq:gmf}) and  corresponding one for the HP monopole (\ref{eq:6}).

As a result of the symmetry breaking, the space-time, for $r \gg \delta $, differs from the standard Schwarzschild metric corresponding to a massive object with mass $M_{\rm core}$ (assuming that all the mass of the monopole is concentrated in the core's interior):
\begin{equation}\label{asympt}
ds^2  = \Big( 1 - 8\pi \, G_{\rm N} \eta^2 - \frac{2 G_{N}\, M_{\rm core}}{r} \Big) dt^2 - \frac{dr^2}{1 - 8\pi \, G_{N} \eta^2 - \frac{2 G_{\rm N}\, M_{\rm core}}{r}} + r^2 \Big( d\theta^2 + {\rm sin}^2 \theta \, d\phi ^2 \Big)~, \quad r \gg \delta~,
\end{equation}
where $(r,\theta,\phi)$ are spherical polar coordinates. The Schwarzschild metric is obtained in the unbroken phase ( $\eta \to 0$). In the asymptotic limit $ r \to \infty$, upon appropriate rescaling of the time $t \to (1 - 8\pi \, G_{N} \eta^2)^{-1/2} \, t^\prime $, and radial coordinate $r$, $r \to (1 - 8\pi \, G_{N} \eta^2)^{1/2}\, r^\prime$, the space-time (\ref{asympt}) becomes a Minkowski space-time with a conical \emph{deficit solid angle} $\Delta \Omega = 8\pi G_{N} \, \eta^2$: 
\begin{equation}\label{asymptflat}
ds^2  = d{t^\prime}^2 - d{r^\prime}^2 -   \Big( 1 - 8\pi \, G_{N} \eta^2 \Big)\, {r^\prime}^2 \Big( d\theta^2 + {\rm sin}^2 \theta \, d\phi ^2 \Big)~, \quad r \gg \delta~. 
\end{equation}
The space-time (\ref{asymptflat}) is not flat, since the scalar curvature behaves as $R \propto 16\pi \, G_N \, \eta^2/r^2$. 
The presence of such a monopole-induced deficit solid angle, can have important physical consequences for scattering processes in such space-times: the scattering amplitude in the forward direction is very large~\cite{papavassiliou} in 
angular regions of order of the deficit angle (or equivalently the squared ratio of the monopole mass to the Planck mass). 

After the initial work of \cite{vilenkin}, a debate has taken place regarding the stability of the configuration~\cite{debate}, which is still ongoing; we shall comment on this debate briefly at the end of our article. Subsequent to the work of \cite{vilenkin} more detailed analysis of the gravitational back reaction effects of such defects has been performed, by requiring a matching of the solutions of the non-linear coupled system of gravitational and matter equations at the core radius; thus the core size is determined dynamically, rather than heuristically from flat space arguments as in the work of \cite{vilenkin}. 
Indeed, in \cite{negative}, the core radius $r_c = 2 \, \lambda^{-1/2}\, \eta^{-1}$ for the self-gravitating solution was found by matching an exterior  Schwarzschild-like metric 
$$ ds^2 = \Big(1 - 8\pi G_{N}\, \eta^2 \, - \frac{2\, G_N\, M}{r} \Big) dt^2 - \Big(1 - 8\pi G_{N}\, \eta^2 \, - \frac{2\, G_N\, M}{r} \Big)^{-1}
 dr^2 - r^2 \, d\Omega^2~, $$ to an interior local de Sitter metric 
$$ ds^2 = \Big(1 - \mathcal{H}^2 \, r^2 \Big) dt^2 - \Big(1 - \mathcal{H}^2 \, r^2 \Big)^{-1} dr^2 - r^2 \, d\Omega^2~ $$
where $M$ denotes the monopole mass and $\mathcal{H}^2 = \frac{8\pi G_{N}\, \lambda \, \eta^4}{12}$ the de Sitter parameter.
Here $\eta$ denotes a quantity with dimension of mass.   Unfortunately such a matching yields a negative mass for the monopole, $M\sim -6\pi \lambda^{-1/2} \eta <0$~\footnote{The motivation to use such a matching comes from the observation that at the origin ($r \to 0$) the Higgs potential for the scalars leads to a cosmological constant $\propto \eta^4$, since any ``matter'' scalar  fields go to zero. However, if a black hole or other geometric singularity is present as $r \to 0$, like in our case with an induced Reissner-Nordstrom geometry due to the antisymmetric tensor and electromagnetic fields (to be discussed below),  the space-time for small $r$ ($r \to 0$)  is different. The argument leading to negative mass would not then hold.}. The interpretation of this sign in~\cite{negative} is based on  the repulsive nature of gravity induced by the vacuum-energy $H^2$ provided by the global monopole. Moreover it has been argued  \cite{negative} that this interpretation is consistent with the monopole being an entity with complicated structure rather than an elementary particle-like excitation. Such a construction with negative mass would not be of relevance to collider physics~\footnote{A classification of the space-times arising from a self-gravitating global monopole solution of the type considered in \cite{vilenkin} and in \cite{negative}, \emph{i.e.} in field theories with only the triplet of the Higgs-type scalar fields and the Ricci scalar curvature, has been given in \cite{bronnikov}, where it was argued that, upon requiring \emph{regularity at the centre} of the monopole, but otherwise independently of the shape of the Higgs potential, the metric can contain at most one horizon, and, in case there is an horizon, the global space-time structure is that of a de Sitter space-time.}.

 As compared to the model for the global monopole, our model (see the next section \ref{sec:3}) includes additional fields, which allow for a positive mass solution, albeit from a `bag model' standpoint. 
Our model contains the (non-gauged) scalar field triplet $\chi^{a}$ of the global monopole 
model, an Abelian $U(1)$ gauge field of electromagnetism with Maxwell tensor $f_{\mu\nu}$,  an extra $O(3)$ singlet scalar field and the tensor $H_{\mu\nu\rho}$ (the field
strength of a $2-$form $B_{\mu\nu}$, the antisymmetric tensor (Kalb-Ramond) field of spin 1). The Maxwell tensor $f_{\mu\nu}$ has a structure similar to (\ref{eq:10}); however, in our case, as we shall discuss later, the first term on the right-hand-side of (\ref{eq:10}) is absent, since we do not have $SU(2)$ gauge fields. The  second term will involve the scalar triplet field $\chi^{a}$, as well as
the $O(3)$ singlet scalar field (either a constant dilaton  or an ultraheavy scalar), stabilised to a constant value (e.g. the minimum value of a scalar potential).

\section{The Model and its Background \label{sec:3}}

In this section we discuss our model for the magnetic monopole and the analytic form of its asymptotic solutions, for large and small distances from the monopole core. We will first describe the Lagrangian of the model, which may be viewed either as purely phenomenological  or as inspired by the bosonic sector of closed string theories upon compactification to four large target-space-time dimensions.The Kalb-Ramond antisymmetric tensor field strength will  determine the magnetic charge of the monopole solution~\cite{Hammond:2002rm}. In four dimensions the Kalb-Ramond field is equivalent to a massless pseudoscalar (gravitational axion-like) field $b(x)$~\cite{aben}. 

\subsection{A Model for a Self-gravitating Global Monopole with Kalb-Ramond Torsion}

Our model is given by the effective 4-dimensional
Lagrangian density $L$ involving the graviton $g_{\mu\nu}$, the
antisymmetric Kalb-Ramond field $B_{\mu\nu}$, the electromagnetic
field tensor $f_{\mu\nu}$,  a real scalar field $\Phi$, whose origin and importance will be discussed in detail below, and the triplet Higgs-like scalar $\chi^{a}$. The latter is associated with the spontaneous breaking of a global $O(3)$ group down to a global $O(2)$. The Goldstone theorem implies the existence of massless 
Goldstone Bosons in such a case, which will be neutral under the Standard Model group. As we shall discuss later, our monopole solutions are expected~\cite{vilenkin} to lose energy and annihilate (with their antimonopoles) through such Goldstone radiation. The Lagrangian density reads:
\begin{eqnarray}
L&=&\left(-g\right)^{1/2}\Big\{ \frac{1}{2}\partial_{\mu}\chi^{a}\partial^{\mu}\chi^{a}-\frac{\lambda}{4}\left(\chi^{a}\chi^{a}-\eta^{2}\right)^{2} -R \nonumber \\
 &+& \frac{1}{2}\partial_{\mu}\Phi\partial^{\mu}\Phi-V\left(\Phi\right) -\frac{1}{12}\, e^{-2\gamma \Phi}\, H_{\rho\mu\nu}H^{\varrho\mu\nu}-\frac{1}{4}\, e^{-\gamma \Phi}\, f_{\mu\nu}f^{\mu\nu}\Big\} \label{eq:13} \end{eqnarray}
where $\gamma$ is a real constant, $g=\det\left(g_{\mu\nu}\right)$, $R$ is the Ricci scalar for
$g_{\mu\nu}$, and the antisymmetric tensor field strength $H_{\rho\mu\nu}=\partial_{\left[\rho\right.}B_{\mu\left.\nu\right]}$, where the brackets $[\dots]$ denote total antisymmetrization of the respective indices. The quantity $\eta > 0$ plays the role of the vacuum expectation value of the Higgs-field in the symmetry broken phase. We shall assume that a singular gauge field $A_{\mu}$ (up to a gauge transformation) may be associated with $f_{\mu\nu}$, on using  a construction outlined by Halpern~\cite{halpern}. 

 In the case of string-inspired models~\cite{Hammond:2002rm,string}, the constant $\gamma = 1$. In such a case $\Phi$ is the dilaton field of the massless string multiplet, and $V(\Phi)$ is a dilaton potential, possibly generated by string loops - the dilaton potential is absent at tree-level in string theory. In addition to string theory, we shall also consider another version of the model, in which $\gamma = 0$. In such a case the field $\Phi$ may be a superheavy real scalar field that is stabilised by its potential $V(\Phi)$ to be some constant value. 
We also assume that, once the scalar field (or dilaton) is stabilised, its potential vanishes (similar to the case of a Higgs-like potential), so there are no contributions to the stress tensor. 
We shall see that the presence of the extra scalar degree of freedom in either case is essential for the association of the ``Kalb-Ramond torsion charge'' with the magnetic charge of the monopole. Notice that in our model the $\chi^a$-matter in the Einstein frame is assumed to be decoupled from the dilaton~\footnote{In the context of string theory effective actions, this can be achieved as follows: one starts from the $\sigma$-model-frame effective action for the scalar triplet, which has the form:
$$
\int d^4 e^{-\Phi}\, \sqrt{-G^{\rm S}}\, \Big[ \dots + \frac{1}{2}\, \partial^\mu \chi^a \partial_\mu \chi^a - \frac{1}{4}\,\tilde \lambda (\Phi) \, \lambda \Big(\chi^a \, \chi^a - \eta^2\Big)^2 \Big]~, 
$$
where $\dots$ denote the rest of the fields, $\tilde \lambda (\Phi)$ is an appropriate function of the dilaton, to be determined, and $e^{\Phi/2} $ is the string coupling in our normalisation. 
The above form of the action is a standard one in a tree-level string theory model, propagating on a closed spherical world sheet, with the overall factor $e^{-\Phi}$ indicating precisely the appropriate power of the (inverse) string coupling pertinent to this genus two world-sheet surface;
$G^{\rm S}$  denotes the determinant of the $\sigma$--model-frame metric of the space-time which is related to the Einstein-frame metric, $g_{\mu\nu}$, by $ G^{\rm S}_{\mu\nu} = e^\phi \, g_{\mu\nu}$. Passing to the Einstein frame and choosing the function $\tilde \lambda (\Phi) = e^{-\Phi}$, defines the scalar sector self-interaction in such a way that the self-coupling is strong for weak string couplings, we obtain the decoupling of the scalar-triplet-$\chi$ sector from the dilaton in (\ref{eq:13}).}
 or heavy scalar. 

Let us first proceed with the $\gamma=1$ (string) case. 
The Lagrangian (\ref{eq:13}) is in the Einstein frame~\cite{string,aben}, where the Einstein-Hilbert curvature term $R$ in the action is canonically normalised. 
Leaving aside, for the moment, the dilaton equation of motion, the equations of motion for the remaining fields are deduced
from (\ref{eq:13}):
\begin{equation}
g^{\nu\beta}\chi_{,\nu\beta}^{a}+\frac{1}{\sqrt{-g}}\partial_{\nu}\left(\sqrt{-g}g^{^{\nu\beta}}\right)\chi_{,\beta}^{a}=-\lambda\eta^{2}\left(\chi^{b}\chi^{b}-\eta^{2}\right)\chi^{a},\label{eq:14}
\end{equation}

\begin{equation}
\nabla_{\kappa}\Big(e^{-2\gamma \Phi}\, H^{\kappa\beta\gamma}\Big)=0,\label{eq:15}
\end{equation}

\begin{equation}
\nabla_{\lambda}\Big(e^{-\gamma \Phi}\, f^{\lambda\kappa}\Big)=0,\label{eq:16}
\end{equation}
and 
\begin{equation}
G_{\mu\nu}=g_{N}\Theta_{\mu\nu}\label{eq:17}
\end{equation}
 where $G_{\mu\varrho}$ is the Einstein tensor , $\Theta_{\mu\nu}$
 is the energy-momentum tensor and 
\bea\label{gndef}
g_{N}=8\pi G_{N}~,
\eea
where $G_{N} = 1/M_P^2$ 
is Newton's constant, with $M_P$ the Planck mass. These equations are supplemented with the Bianchi
identity for the Kalb-Ramond field strength, stemming from its definition: 
\begin{equation}
\epsilon^{\mu\nu\lambda\rho}\partial_{\rho}H_{\mu\nu\lambda}=0.\label{eq:18}
\end{equation}
Furthermore in $4-$dimensions the Kalb-Ramond field strength is dual
to a pseudoscalar (``axion''-like) field $b$~\footnote{ In string theories~\cite{string}, the field strength
$H_{\mu\nu\rho}$, in the presence of gauge fields $A_{\mu}$, is no longer given only by the curl of $B_{\mu\nu}$ but contains additional parts proportional to the Chern-Simons three form $A \wedge F $. Such terms lead to higher derivative terms in the string effective action, and are ignored in our model. Their inclusion for Abelian gauge fields could lead to additional interesting electromagnetic effects~\cite{cherneff}, which, however, are not of interest to us here. }:
\begin{equation}
H_{\mu\nu\lambda}= e^{2\Phi}\, \epsilon_{\mu\nu\lambda}^{\;\;\;\;\:\;\sigma}\partial_{\sigma}b ~, \label{eq:21}
\end{equation}
where 
\bea\label{lct}
\tilde \epsilon_{\mu\nu\rho\sigma} = \sqrt{-g} \, \epsilon_{\mu\nu\rho\sigma}~,
\eea 
is the flat space-time Levi-Civita symbol, $ \epsilon_{\mu\nu\rho\sigma}$ is the covariant Levi-Civita tensor density, with $\tilde \epsilon_{\mu\nu\rho\sigma}$  $\tilde \epsilon_{0123} = +1$ {\it etc.} (and also $\epsilon^{\mu\nu\rho\sigma} = \sqrt{-g} \, \tilde \epsilon^{\mu\nu\rho\sigma}$). The form (\ref{eq:21}) for the field strength satisfies (\ref{eq:15}) automatically, taking into account that the gravitational covariant derivative is defined in terms of the usual symmetric Christoffel symbol.

It is important to note that
 in our approach we shall concentrate on the \emph{dual theory}, where the physical degree of freedom for the Kalb-Ramond field is the axion $b(x)$, defined in (\ref{eq:21})~\footnote{In a Feynman path-integral formulation the dual theory for the Lagrangian (\ref{eq:13}) (in a Minkowski-signature space-time) is obtained~\cite{fermi2,fermi} by implementing the Bianchi constraint  (\ref{eq:18}) via a path-integral $\delta$-function, which can then be represented using a Lagrange multiplier field $b(x)$. The dual theory is obtained on integrating out the field $H_{\mu\nu\rho}$  in the path-integral. In the language of differential forms, the latter constraint reads $\,\textbf{d}^*\textbf{S} = 0$, where $\textbf{S} = {}^\star\,\textbf{H}$  is the dual form of the Kalb-Ramond field strength \textbf{H}: $S_d
=   \frac{1}{3!}     \epsilon^{abc}_{\quad   d}   H_{abc}$. Imposing this constraint on the full quantum theory, 
is equivalent to imposing an exact conservation of the ``Kalb-Ramond-torsion charge'' $Q = \int \,^*S = 0$. We then have in a path integral:
\begin{eqnarray}
 \label{qtorsion}
{\mathcal Z} &\propto& \int D \textbf{S} \, \exp \Big[ i \int  \frac{3e^{-2\Phi}}{4 g_N} \textbf{S} \wedge {}^\star\! \textbf{S} \Big]\, \delta \Big(\textbf{d} {}^\star\! \textbf{S}\Big)\, 
= \, \int D \textbf{S} \, D b   \, \exp \Big[ i \int
    \frac{3e^{-2\Phi}}{4g_N} \textbf{S} \wedge {}^\star\! \textbf{S} +
      \Big(\frac{3}{2g_N}\Big)^{1/2} \, b \, \textbf{d} {}^\star\! \textbf{S}
      \Big] \nonumber \\
      & \propto & \int  D b  \, \exp\Big[ -i \int \frac{1}{2}e^{2\Phi}\,
      \textbf{d} b\wedge {}^\star\! \textbf{d} b  \Big]~,
\end{eqnarray}
where the various proportionality factors represent appropriate normalisations of the various forms, and we work with a dimensionful field $b(x)$ with mass-dimension one; above we wrote explicitly only the part of the quantum path integral of the Lagrangian (\ref{eq:13}) that involves the (dual of the) Kalb-Ramond field $\textbf{S}$, which is relevant for our discussion here. The  ``non-propagating''   $\textbf{S}$  field  has  been  integrated out completely, upon implementing the Bianchi constraint (\ref{eq:18}), and partially integrating the second term in the argument of the exponential in the middle equation of (\ref{qtorsion}). We note  the change in sign of the kinetic term of $b$ as compared with that of $\textbf{S}$, and the different scalings with the dilaton $\Phi$ between these two terms. This results in a path-integral over the pertinent Lagrange multiplier field $b$ that leads to the equations of motion (\ref{eq:dil}).}. 
In the context of the dual theory, when one considers the dilaton equations of motion from the Lagrangian (\ref{eq:13}) with $\gamma=1$, one has to take into account the non-trivial variation  
$\delta H_{\mu\nu\rho} /\delta \Phi = 2 H_{\mu\nu\rho} = 2 e^{2\Phi} \epsilon_{\mu\nu\rho\sigma}\, \partial^\sigma b $. With this in mind, it is then straightforward to see that the dilaton equation of motion obtained from the Lagrangian (\ref{eq:13}) implies:
\begin{equation}\label{eq:dil}
e^{2\Phi}\partial_\mu b \, \partial^\mu b +  \frac{1}{4}e^{-\Phi} \, f_{\mu\nu}\, f^{\mu\nu} - \frac{\delta V(\Phi)}{\delta \Phi} +  {\mathcal O}(\partial \Phi) = 0~,
\end{equation}
where we did not write explicitly the terms involving $\partial_\mu \Phi$, since we will be interested in situations in which the dilaton is stabilised to a constant value $\Phi=\Phi_0$,
which may occur at the minimum of its potential when
\begin{equation}\label{eq:dilpotmin}
\frac{\partial V(\Phi)}{\partial \Phi }\Big|_{\Phi = \Phi_0} = 0~, \quad {\rm with} \quad V(\Phi_0) = 0 ~.
\end{equation}
Thus, for a constant dilaton, the case of interest, the dilaton equation (\ref{eq:dil}) implies a constraint on the Kalb-Ramond and Maxwell field strengths. This constraint will be at the heart of our considerations later on in the article, when we link the Kalb-Ramond torsion charge with the magnetic charge of the electromagnetic monopole.

In the limiting non-stringy case $\gamma=0$, we may ensure the stabilization of the heavy scalar field  to a constant value $\Phi=\Phi_0$ by imposing again (\ref{eq:dilpotmin}); however in this limiting case the constraint (\ref{eq:dil}) between the Kalb-Ramond field strength and the Maxwell tensor  is not imposed. Nevertheless, even in this case, we shall see that an appropriate modification of the Maxwell tensor in the spirit of (\ref{eq:10}), involving the $H_{\mu\nu\rho}$ field and the heavy scalar $\Phi$, can be constructed which remarkably still solves Maxwell's equation (\ref{eq:16}) for $\Phi=\Phi_0=$ constant. We next proceed to solving  the equations (\ref{eq:14}), (\ref{eq:15}), (\ref{eq:16}) and (\ref{eq:dil}) (with the condition (\ref{eq:dilpotmin})).

\subsection{Solution of the Model Equations: Ans\"atze}

We will consider static solutions of the equations (\ref{eq:14}), (\ref{eq:15}), (\ref{eq:16}), (\ref{eq:17}),
and (\ref{eq:18}) by making the ans\"atze
\begin{equation}
g_{\mu\nu}=\left(\begin{array}{cccc}
B(r)\\
 & -A(r)\\
 &  & -r^{2}\\
 &  &  & -r^{2}\sin^{2}\theta
\end{array}\right)\label{eq:19}
\end{equation}
and 
\begin{equation}
f_{\mu\nu}=\left(\begin{array}{cccc}
0 & 0 & 0 & 0\\
0 & 0 & 0 & 0\\
0 & 0 & 0 & 2r\sin\theta\, W(r)\\
0 & 0 & -2r\sin\theta\, W(r) & 0
\end{array}\right).\label{eq:20}
\end{equation}
This ansatz for $f_{\mu\nu}$ is compatible with the ansatz
for $f_{\mu\nu}$ in the HP solution and satisfies (\ref{eq:16}).
The associated magnetic field has only a radial component, which in contravariant form reads:
\bea\label{magnetic}
\mathcal{B}^r = \epsilon^{r \theta \phi}\, f_{\theta\phi} = \frac{1}{\sqrt{-g}} \, \eta^{r \theta \phi}\, f_{\theta\phi} = 2 \, \frac{W(r)}{r}~,
\eea
where $\eta^{r \theta \phi}=+1$, \emph{etc} is the three-space totally antisymmetric symbol, and we took into account Eq.~(\ref{eq:19}). The electric field is zero.

The ansatz for the scalar field is 
\begin{equation}
\chi^{a}=\eta f\left(r\right)\frac{x^{a}}{r}.\label{eq:22}
\end{equation}
The ans\"atze in (\ref{eq:19}) and (\ref{eq:22}) are those for the gravitational
monopole. Since in addition we have the electromagnetic and Kalb-Ramond
tensors, we can investigate whether
the gravitational monopole induces a magnetic monopole from the enlarged set of equations.

On using the ans\"atze in the Einstein equation (\ref{eq:17}),
we obtain 
\bea
\frac{-A\left(r\right)+A^{2}\left(r\right)+rA'\left(r\right)}{g_{N}A^{2}\left(r\right)} &=& 2W^{2}\left(r\right)+\frac{1}{4}b'^{2}\left(r\right)\frac{r^{2}}{A\left(r\right)}+\frac{\eta^{2}}{2}\left(2f^{2}\left(r\right)+\frac{f\,'^{2}\left(r\right)r^{2}}{A\left(r\right)}\right) \nonumber \\
&+&  \frac{\lambda}{4}\eta^{4}\left(f^{2}\left(r\right)-1\right)^{2}r^{2},\label{eq:23}
\eea
\bea
\frac{B\left(r\right)-A\left(r\right)B\left(r\right)+rB'\left(r\right)}{g_{N}B\left(r\right)}&=&-2A\left(r\right)W^{2}\left(r\right)+\frac{1}{4}b'^{2}\left(r\right)r^{2}+\frac{1}{2}\eta^{2}r^{2}f\,'^{2}\left(r\right)  \nonumber \\ 
&-&\eta^{2}A\left(r\right)f^{2}\left(r\right)- \frac{\lambda}{4}\eta^{4}A\left(r\right)\left(f^{2}\left(r\right)-1\right)^{2}r^{2},\label{eq:24}
\eea
and
\bea
&& \frac{r}{4g_{N}}\left(2\frac{A'\left(r\right)}{A\left(r\right)}+\frac{rB'^{2}\left(r\right)}{B^{2}\left(r\right)}+\frac{rA'\left(r\right)B'\left(r\right)}{A\left(r\right)B\left(r\right)}-2\left(\frac{B'\left(r\right)}{B\left(r\right)}+\frac{rB''\left(r\right)}{B\left(r\right)}\right)\right)
\nonumber \\
&=&-2W^{2}\left(r\right)A\left(r\right)+\frac{1}{4}r^{2}b'^{2}\left(r\right)+\frac{r^{2}\eta^{2}}{2}f\,'^{2}\left(r\right)+\frac{\lambda\eta^{4}}{4}A\left(r\right)r^{2}\left(f^{2}\left(r\right)-1\right)^{2}, \nonumber \\ \label{eq:25}
\eea
where prime indicates derivative with respect to $r$. 
Furthermore (\ref{eq:14}) leads to 
\begin{equation}
\frac{f\,''\left(r\right)}{A\left(r\right)}-\frac{1}{2A\left(r\right)}\left(\frac{A'\left(r\right)}{A\left(r\right)}-\frac{B'\left(r\right)}{B\left(r\right)}-\frac{4}{r}\right)f^\prime\, 
\left(r\right)-\frac{2f\left(r\right)}{r^{2}}=\lambda\eta^{2}\left(f^{2}\left(r\right)-1\right)f\left(r\right).\label{eq:26}
\end{equation}
The last remaining equation, derived from (\ref{eq:18}), is 
\begin{equation}
\frac{d}{dr}\left(\sqrt{\frac{B\left(r\right)}{A\left(r\right)}}r^{2}\frac{db}{dr}\right)=0.\label{eq:27}
\end{equation}
Its solution is

\begin{equation}
b'\left(r\right)=\frac{\varsigma}{r^{2}}\sqrt{\frac{A\left(r\right)}{B\left(r\right)}}\label{eq:28}
\end{equation}
where $\varsigma$ is a constant of integration which measures the strength
of the Kalb-Ramond field strength.  

It is necessary to be aware of units of variables and
so we recast the equations in terms of dimensionless variables: 
\bea\label{scaled}
W\rightarrow\frac{W}{\sqrt{g_{N}}},\quad r\rightarrow\sqrt{g_{N}}\, r,\quad b\rightarrow\frac{b}{\sqrt{g_{N}}},\quad \eta\rightarrow\frac{\eta}{\sqrt{g_{N}}}.
\eea
The equations satisfied by these rescaled variables are the same as
(\ref{eq:23}), (\ref{eq:24}) and (\ref{eq:25}) but with $g_{N}$ replaced
by $1$.

\subsection{Analytical Solution of the Model equations: Asymptotic analysis}

Equations (\ref{eq:23}), (\ref{eq:24}), (\ref{eq:25}), 
and (\ref{eq:26}) will be solved in two asymptotic regions, the
near-field ($r\rightarrow0$) and far-field ($r\rightarrow\infty$). The existence of the full interpolating solution then is assumed and based on continuity in space.
Approximate interpolating solutions will be discussed in a subsequent paper. 
In both regions,  \emph{to leading order}, we will require 
\bea\label{eq:AB}
B\left(r\right) \simeq A^{-1}\left(r\right)~,
\eea
which is certainly required for the far field (Newtonian) limit. However, as we shall show, the presence of a non-trivial antisymmetric tensor field strength (\ref{eq:28}) and of the scalar triplet field with non-trivial vacuum expectation value $\eta$, imply (next-to-leading order) modifications in (\ref{eq:AB}), which are \emph{crucial} for the consistency of the solutions. In particular, as we shall discuss below, for the small $r$ ($r \to 0$) region, we find
\bea\label{eq:ABsmallr}
B\left(r\right) \, A\left(r\right) = 1 + \mathcal{O}(r^2)~, \quad r \to 0~, 
\eea
while for the large $r \to \infty$ region we have:
\bea\label{eq:ABlarger}
B\left(r\right) \, A\left(r\right) = 1 + \mathcal{O}(\frac{1}{r^2})~, \quad r \to \infty~.
\eea
Working in units with $g_N=1$, the necessity of such deviations in both regions can be seen by manipulating eqs. (\ref{eq:23}) and (\ref{eq:24}) to rewrite them as:
\bea
1 - \frac{1}{A} + \frac{rA'\left(r\right)}{A^{2}\left(r\right)} &=& 2W^{2}\left(r\right)+\frac{1}{4}b'^{2}\left(r\right)\frac{r^{2}}{A\left(r\right)}+\frac{\eta^{2}}{2}\left(2f^{2}\left(r\right)+\frac{f\,'^{2}\left(r\right)r^{2}}{A\left(r\right)}\right) \nonumber \\
&+&  \frac{\lambda}{4}\eta^{4}\left(f^{2}\left(r\right)-1\right)^{2}r^{2}\label{eq:23b}
\eea
and
\bea
1 - \frac{1}{A} - \frac{rB'\left(r\right)}{A \, B} &=& 2W^{2}\left(r\right)- \frac{1}{4}b'^{2}\left(r\right)\frac{r^{2}}{A\left(r\right)}+\frac{\eta^{2}}{2}\left(2f^{2}\left(r\right)-\frac{f\,'^{2}\left(r\right)r^{2}}{A\left(r\right)}\right) \nonumber \\
&+&  \frac{\lambda}{4}\eta^{4}\left(f^{2}\left(r\right)-1\right)^{2}r^{2}. \label{eq:24b}
\eea
If (\ref{eq:AB}) were to hold exactly, then one would have 
$\frac{A^\prime}{A} = -\frac{B^\prime}{B}$, which would make the left hand sides of (\ref{eq:23b}) and (\ref{eq:24b}) identical and, on 
subtracting the equations, it would yield
\bea\label{eq:inc}
0 = \frac{r^2}{A} \Big(\frac{1}{4} (b^\prime)^2 + (f^\prime)^2  \Big). 
\eea
As we shall discuss below, for small $r\to 0$ one has $B \sim p_0/r^2$, $p_0 > 0$ a constant, and $f^\prime = f_0 $=constant (\emph{cf.} (\ref{eq:f0}));  if  (\ref{eq:inc}) had been valid then we would have $\varsigma=f_0=0$. 
For large $r$, the ansatz we take for the function $f(r)$ (\emph{cf.} 
(\ref{core}),(\ref{epscore}) for $r \to \infty$) makes the contributions of the $f^\prime$ -terms in (\ref{eq:23b}) and (\ref{eq:24b}) subleading in the region $r \to \infty$, as compared to the rest of the terms. Upon ignoring such terms then, and subtracting the latter two equations leads to $\varsigma =0$. 

The leading order assumption (\ref{eq:AB}) is used in many non-trivial black-hole solutions, e.g. for the Reisser-Nordstr\"om (RN) black  hole solution  the metric is~\cite{RN}:
\begin{equation}
\label{RN}
ds^{2}=\Delta dt^{2}- \Delta^{-1}dr^{2}-r^{2}d\Omega^{2}
\end{equation}
where $d\Omega^{2}$ is the metric on a $2$-sphere and $\Delta=1-\frac{2G_NM}{r}+\frac{G_N\, \mu^{2}}{r^{2}}$ with $\mu$ being the magnetic charge, and $M$ the mass of the black hole. Consequently the assumption of $B\left(r\right)=A^{-1}\left(r\right)$ is a useful one. The RN black hole is not singular at the horizons, the apparent singularities being co-ordinate artefacts.\footnote{In Maxwell-Einstein systems the effects of ordinary axion fields (which differ from those associated with our three form $H_{\mu\nu\rho}$)  have been discussed in \cite{lee}, with the conclusion that the axion charge, which we identify with $-\varsigma$ (see discussion below, 
Eq.~(\ref{baxion})), contributes to the charge-terms in a metric pertaining to a RN black hole; in string theories with antisymmetric tensor fields present, there are rotating black hole solutions of Kerr-Newmann-Reissner-Nordstrom type~\cite{sen}; charged non-rotating black hole solutions are present in string-inspired models with dilaton, gauge and Kalb-Ramond axion fields present, but without scalar triplet fields $\vec \chi$, associated with the global monopole~\cite{axionbh}.}  In our case, in view of the corrections (\ref{eq:ABsmallr}), (\ref{eq:ABlarger}), we obtain deformed RN-type solutions, but the shielding of the curvature singularities at $r=0$ by horizons (i.e. absence of naked singularities) holds for sufficiently large mass compared to the charge. (In our case of small masses the naked singularities can still be shielded; see discussion in section \ref{sec:rnmass}, following Eq.~(\ref{rnhor})) .

Because
of the use of scaled \emph{{}dimensionless} variables,
when $r$ is of $O\left(1\right)$ the physical $r$ is order of the
Planck length. At this scale the equations cannot be expected to be
valid because of quantum gravity corrections; hence, in order to be able
to estimate the magnetic energy of our monopole, we put the effective Planck length as a lower
distance cut-off; in the estimate we will use  expressions for our dependent
variables which represent their leading asymptotic behaviour 
for $r\rightarrow0$ and  $r\rightarrow\infty$.

\subsubsection{Small $r$ analysis}

Asymptotically, for small $r$, let us write $B\left(r\right)\sim\frac{p\left(r\right)}{r^{2}}$
and assume that 
\bea\label{eq:f0}
f\left(r\right)\sim f_{0}\, r~,
\eea (which is consistent with the scalar field equation of motion (\ref{eq:26}) in the limit $r \to 0$ and is similar to the $r$ dependence found in the construction of
the HP monopole~\cite{hpmono}). From (\ref{eq:23}) and (\ref{eq:25}) we deduce
that 
\begin{equation}
1-\frac{p''\left(r\right)}{2}=\frac{1}{2}\lambda\eta^{4}r^{2}\left(f_{0}^{2}r^{2}-1\right)^{2}+\frac{\varsigma^{2}}{2p\left(r\right)}+\eta^{2}f_{0}^{2}\left(r^{2}+p\left(r\right)\right).\label{eq:29}
\end{equation}
We cannot solve this equation without approximation; on the right-hand
side of (\ref{eq:29}), in the denominator of the term proportional to $\varsigma^2$,  we consider $p\left(r\right)$  to be approximately
a non-zero constant $p_{0}$ which leads to the equation 
\begin{equation}
1-\frac{p''\left(r\right)}{2}=\frac{1}{2}\lambda\eta^{4}r^{2}\left(f_{0}^{2}r^{2}-1\right)^{2}+\frac{\varsigma^{2}}{2p_{0}}+\eta^{2}f_{0}^{2}\left(r^{2}+p\left(r\right)\right).\label{eq:30}
\end{equation}
The general solution of (\ref{eq:30}) is 
 \bea
p\left(r\right) &=& c_{2}\sin\left(\sqrt{2}\text{\ensuremath{f_{0}}}\eta r\right)+c_{1}\cos\left(\sqrt{2}\text{\ensuremath{f_{0}}}\eta r\right) 
+ \frac{\mathcal{Z}}{2f_{0}^{4}\eta^{4}p_{0}}~, \nonumber \\
 {\mathcal Z} &=& \eta^{2}\left(-\varsigma^{2}f_{0}^{2}-90f_{0}^{2}\lambda p_{0}r^{2}+4f_{0}^{2}p_{0}+12\lambda p_{0}\right) \nonumber \\ &+& \eta^{4}\left(15f_{0}^{4}\lambda p_{0}r^{4}-12f_{0}^{2}\lambda p_{0}r^{2}-2f_{0}^{4}p_{0}r^{2}+\lambda p_{0}\right)  \nonumber \\ &+& \eta^{6}\left(f_{0}^{6}\lambda p_{0}\left(-r^{6}\right)+2f_{0}^{4}\lambda p_{0}r^{4}-f_{0}^{2}\lambda p_{0}r^{2}\right)+90\lambda p_{0}
\label{eq:31}
\eea
where $c_{1}$ and $c_{2}$ are constants of integration. Since $\eta$
is small (on assuming that the symmetry breaking scale is much smaller
than the Planck scale), $p\left(r\right)$ is well approximated by
a constant near $r=0.$ Hence~\footnote{In fact, on assuming (\ref{eq:AB}) it is easily seen that  (\ref{eq:26}) can be written as:
\bea\label{infactor}
\frac{d}{dr} \Big(f^\prime \, r^2 \, B \Big) = 2 f + \lambda \eta^2 (f^2 - 1) \, f\, r^2 
\eea
which can be readily integrated to yield 
\bea\label{Bexpr}
B = \frac{c_0}{f^\prime \, r^2} + \frac{1}{f^\prime \, r^2} \, \int d\tilde r \Big[2f(\tilde r) + \lambda \eta^2 (f(\tilde r)^2 - 1) \, f(\tilde r) \, {\tilde r}^2\Big]~,
\eea
where $c_0$ is an integration constant. 
Upon assuming $f(r) \sim f_0 r $ for $r \to 0$, the small $r$ behaviour deduced from (\ref{Bexpr}), is consistent with (\ref{eq:32}), obtained from our small $r$ analysis, upon fixing the constants $p_0 = c_0/f_0$.  
However, for large $r$, although (\ref{eq:AB}) is assumed, and one might have surmised that (\ref{Bexpr}) would still be valid, this is not the case: 
$f \simeq 1 -\frac{\alpha_1}{r^2}$, with $\alpha_1$ a constant, which implies that as $r \to \infty$, $f^\prime \to 0$ (and that $f^\prime \, r^2 \sim r^{-1}$) .  Hence the derivation of  (\ref{Bexpr}) entails an implicit division by zero, which is an inappropriate operation.}, 
 upon making \emph{the leading order approximation} $B\left(r\right) \simeq A^{-1}(r)$, we find
\begin{equation}
B\left(r\right)  = \frac{p_{0}}{r^{2}}~,  \quad \rm{for}\quad r \to 0.
\label{eq:32}
\end{equation}
We keep this expression for $B(r)$ and now proceed to find the next-to-leading order corrections in the product $A\, B$ which are induced by the presence of the antisymmetric tensor and the non-trivial vacuum expectation value of the scalar fields (i.e. the global monopole).
Let us assume that, for small $r \to 0$:
\bea
A (r) \, B (r) = 1 + \epsilon (r) 
\label{eq:abnot1}
\eea
where $\epsilon (r) \to 0$ as $r \to 0$. This implies 
\bea
\frac{A^\prime}{A} = - \frac{B^\prime}{B}  + \frac{\epsilon^\prime (r)}{1 + \epsilon(r)}~.
\label{eq:abprimes}
\eea
Upon substituting (\ref{eq:32}), (\ref{eq:f0})  and (\ref{eq:abprimes}) into the Einstein equations (\ref{eq:23b}) and (\ref{eq:24b}), and subtracting them,  we obtain, for small $r$, to leading order,
\bea\label{eq:eprime}
\frac{r\, \epsilon^\prime}{1 + \epsilon } \frac{1}{A} = \frac{\varsigma^2}{2 p_0} + \frac{f_0^2 \eta^2 \, p_0  }{1 +\epsilon}.
\eea
Upon assuming $\epsilon (r) = {\mathcal O}(r^2)$, we seek consistent solutions in the region $r \to 0$. In this case, the denominator of the second term on the right hand side of (\ref{eq:eprime}) can be approximated by unity, which on account of (\ref{eq:abnot1}), yields 
\bea\label{eq:3.37}
\frac{\epsilon^\prime}{(1 + \epsilon)^2} = - \frac{d}{dr} \Big(\frac{1}{1 + \epsilon} \Big) = \Big(\frac{\varsigma^2}{2 p_0^2} + f_0^2 \eta^2 \Big)\, r 
\eea
which can be integrated to give 
\bea\label{epsilonr}
\epsilon(r) = \frac{1}{1 - \Big(\frac{\varsigma^2}{4 p_0^2} + \frac{\eta^2 f_0^2}{2}\Big) \, r^2} - 1 = \Big(\frac{\varsigma^2}{4 p_0^2} + \frac{\eta^2 f_0^2}{2}\Big)\, r^2 + \dots ~, \quad r \to 0~, 
\eea
upon imposing the requirement that $\epsilon (0)=0$. 

Although we have used so far the leading order approximation (\ref{eq:32}) for $B$ (as $ r \to 0$), when dealing with eq. (\ref{eq:23b}) we should  make use of the complete Reissner-Nordstrom expression 
\bea
B = 1 - \frac{2{\mathcal M}}{r} + \frac{p_0}{r^2}
\label{completeRNB}
\eea
with ${\mathcal M}$ the monopole mass. Indeed, upon using (\ref{completeRNB}), (\ref{eq:abnot1}), (\ref{eq:abprimes}) and (\ref{eq:3.37}), as $r \to 0$, we obtain:
\bea\label{eq:3.23new}
&& 1 - \frac{B}{1 + \epsilon}  - r \frac{B^\prime }{1 + \epsilon} + \frac{r \, \epsilon^\prime}{(1 + \epsilon)^2} 
\simeq 1 - B + B\, \epsilon - r\, B^\prime + r\, B^\prime \, \epsilon + r \epsilon^\prime \, B\, (1 - 2\epsilon) + \dots  \nonumber \\
&& = \frac{p_0}{r^2}  + \frac{\varsigma^2}{4p_0} + f_0^2 \eta^2 p_0 + \dots = 2 W^2 + \frac{\varsigma^2}{4p_0} + f_0^2 \eta^2 p_0~,
\eea
where the $\dots$ indicate subleading terms that go to zero as $r \to 0$. ${\mathcal M}$ is undetermined since
the  mass terms cancel altogether; the $\varsigma^2$ and $\eta^2 \, f_0^2$ dependent terms also cancel, leaving to leading order (as $ r \to 0$) the relation
\begin{equation}
W^{2}\left(r\right)\sim\frac{p_{0}}{2r^{2}} ~, \, \quad r \ll1~.
\label{eq:34}
\end{equation}

One can easily see that the above results, (\ref {eq:abnot1}) and (\ref{eq:3.37})-(\ref{eq:34}), are also consistent with the third Einstein equation (\ref{eq:25}) in the region $r \to 0$. 
Thus, the constant $p_0 > 0$ cannot be determined in this asymptotic analysis, and the only relation that emerges is (\ref{eq:34}), which was to be expected from the Reissner-Nordstrom character of the metric. 

However, as we shall presently see, the ``charge'' distortion part of the space-time metric function $B$ (\ref{completeRNB}) (\emph{i.e.} the term $p_0/r^2$) comes exclusively from the torsion field: 
\bea
p_0 \propto  \varsigma^2~,
\label{torsionH}
\eea
so that the magnetic charge $g \propto \varsigma $ (\emph{cf.} (\ref{magnetic}) and (\ref{magnfield}), (\ref{magneticcharge}) below). (The normalization factors will be fixed, as we shall see, once we embed the model in a string theory framework.) 
To this end, for the case $\gamma=0$, we define $f_{\mu\nu}$, which is a solution of (\ref{eq:16}), as follows:
\bea\label{hpfmn}
f_{\mu\nu} =   -H_{abc} \phi^a \partial_\mu \phi^b \, \partial_\nu \phi^c.
\eea
The fields $\phi^a$, $a=0,1,2,3$, is a tetrad of the scalar fields, with $\phi^0$= constant, and $\phi^a \, (a=1,2,3)$ identified with the triplet $\chi^a $ (\ref{eq:22}), normalised in such a way that $\phi^a$ which maps the SO(3) internal group onto the sphere $S^3$ of three-space. 
The fourth member  $\phi^0$ of the tetrad is identified with (some function of) the superheavy scalar field $\Phi$ appearing in the Lagrangian (\ref{eq:13}), which is assumed to be stabilised to a constant value at an appropriate minimum of its potential (\ref{eq:dilpotmin}). Using (\ref{eq:21}), (\ref{lct}),
and the fact that the triplet (\ref{eq:22}) defines a $S^3$-spatial coordinate set $(\eta f(r), \theta, \phi)$ which maps the SO(3) internal space to the three space, one has for the non-trivial $\theta\phi$ component of the electromagnetic tensor (\ref{hpfmn}):
\bea\label{ftp}
f_{\theta\phi} = \frac{\phi^0}{A} \, b^\prime \, r^2 \, {\rm sin}\theta = \phi^0 \, \varsigma \, {\rm sin}\theta~.
\eea
(In the last equality we have made use of the exact solution (\ref{eq:28}), a consequence of the Bianchi identity for $H_{\mu\nu\rho}$.) Comparing with 
(\ref{eq:20}) we thus obtain 
\bea\label{wallr}
W(r) = \phi_0\, \frac{\varsigma}{2\, r}
\eea
for all $r$. This is consistent since 
the solution of the Maxwell's equations for the electromagnetic field (\ref{eq:16}) is independent of the explicit form of the function $W(r)$. Eq (\ref{wallr}) 
is also consistent with 
(\ref{torsionH}). 
In view of (\ref{wallr}) and (\ref{magnetic}), the induced magnetic monopole flux is controlled by the  Kalb-Ramond field strength parameter $\varsigma$, \emph{i.e.} a vanishing or constant Kalb-Ramond axion field  leads to the absence of the magnetic monopole. 

 We now remark that in the context of the string-inspired low energy Lagrangian (\ref{eq:13}) with $\gamma = 1$, we may arrive at a similar conclusion 
and moreover we can fix the proportionality coefficient $\phi_0$ in (\ref{wallr}).  Indeed, to this end we first note that, on using (\ref{eq:dilpotmin}), (\ref{eq:28}), as well as the fact that the spatial part of the Maxwell tensor $\frac{1}{4}\, f_{ij}f^{ij} = \frac{1}{2}\, ({\mathcal B}^r)^2 A $, with ${\mathcal B}^r$ 
the radial (and, in our case, the only non-trivial ) component of the magnetic field (\ref{magnetic}), the dilaton equation (\ref{eq:dil}) yields the constraint (for constant dilaton $\Phi=\Phi_0$, which without loss of generality we can take it to be $\Phi_0=0$)
\begin{equation}\label{brnew}
\frac{\varsigma^2}{r^4}\, \frac{1}{B} =  \frac{1}{2}\, ({\mathcal B}^r)^2 \, A \, \Rightarrow \, {\mathcal B}^r = \sqrt{2} \, \, \frac{1}{\sqrt{AB }}\, \frac{\varsigma}{r^2}~,
\end{equation}
The reader should notice that equation (\ref{brnew}) is valid for all $r$. 

On taking into account (\ref{eq:ABsmallr}), \emph{i.e.} that to leading order as $r \to 0$ one has $A B \sim 1$, we see that (\ref{brnew}) implies a singularity structure (as $r \to 0$) for the radial component of the magnetic field of magnetic monopole type~\cite{diracmono}, 
\begin{equation}\label{magmon}
{\mathcal B}^r \simeq \sqrt{2} \, \frac{\varsigma}{r^2} = \frac{g}{r^2}~, 
\end{equation}
with a magnetic charge $g$ being given by  
\begin{equation}\label{magcharge}
g = \sqrt{2} \, \varsigma~,  
\end{equation}
thus fixing the proportionality coefficient in (\ref{torsionH}) to: 
\begin{equation}\label{torsionH2}
p_0 = 2\, \varsigma^2~.
\end{equation}
Thus, from (\ref{eq:34}) we have
\begin{equation}
W(r) = \frac{\varsigma}{r}~.
\label{wsigma}
\end{equation}

We will now examine (\ref{eq:26})
to investigate the consistency of our assumption that $f\left(r\right)\sim f_{0}r$. On
substituting the expression (\ref{eq:32}) for $B\left(r\right)$  in a linearised
form of (\ref{eq:26}), we obtain 
\begin{equation}
f\,''\left(r\right)=\frac{2- \lambda\eta^{2}r^{2}}{p_{0}}f\left(r\right).\label{eq:35}
\end{equation}
These equations can be solved in terms of parabolic cylinder functions
which are analytic in the neighbourhood of $r=0$. A solution exists
which (for small $r$) is proportional to $r+\frac{r^{3}}{3p_{0}}$
and so we have consistent ans\"atze.

\subsubsection{Large r analysis:}

 For large $r$, since we expect the Newtonian limit to hold, we will
consider the ans\"atze
\bea\label{larger}
A\left(r\right)B\left(r\right) = 1 + \frac{\epsilon_0}{r^2}, \, \epsilon_0 \in {\mathbb{R}}, \, 
\quad {\rm and } \quad B\left(r\right)\sim1+\beta_{1}+\frac{\beta_{2}}{r}+\frac{\beta_{3}}{r^{2}}~, \, r \to \infty~,
\eea
which imply
\bea\label{largerab}
\frac{A^\prime}{A} = - \frac{B^\prime}{B} - \frac{2\epsilon_0}{r^3}~, \quad r \to \infty~.
\eea
Also the asymptotic ansatz 
\bea\label{core}
f\left(r\right)=1-\frac{\alpha_{1}}{r^{2}}+\delta\left(r\right),
\eea
where $\alpha_{1}$ is a constant, solves the scalar field equation (\ref{eq:14}). From (\ref{eq:26}), on considering
the leading order in $\frac{1}{r}$,  we find $\alpha_{1}=\frac{1}{\lambda\eta^{2}}$.
From (\ref{eq:23}) and the leading behaviour in $\frac{1}{r}$ we have
the requirement that $\beta_{1}=-\eta^{2}$ which gives the deficit
angle already noted by the authors of ref.~\cite{vilenkin} (using a different argument).
This also matches the leading behaviour in (\ref{eq:24}).  Ignoring, for the moment, the $1/r^2$ corrections on the right-hand-side of the AB product in (\ref{larger}), we observe that the
equation linear in $\delta\left(r\right)$ that is derived from (\ref{eq:26})
is 
\begin{equation}
\left(1-\eta^{2}\right)\frac{d^{2}}{dr^{2}}\delta\left(r\right)+\frac{2}{r}\left(1-\eta^{2}\right)\frac{d}{dr}\delta\left(r\right)-2\lambda\eta^{2}\delta\left(r\right)=0.\label{eq:36}
\end{equation}
This has a decaying solution 
\bea\label{epscore}
\delta\left(r\right) = \frac{\exp\left(-\eta\sqrt{\frac{2\lambda}{1-\eta^{2}}}r\right)}{r}~,
\eea
and so  $\delta\left(r\right)$  in (\ref{core}) is exponentially
small and can be ignored.

On subtracting (\ref{eq:24b}) from (\ref{eq:23b}), and taking into account (\ref{larger}) and (\ref{largerab}),
 we readily obtain, to leading order in $r \to \infty$:
 \bea\label{epsilon0zeta}
 \epsilon_0 = - \frac{\varsigma^2}{4(1 - \eta^2)^2} ~.
\eea
Upon adding (\ref{eq:23b}) and (\ref{eq:24b}) we can determine $W^2$ in the large $r$ region:
\bea\label{wlr2}
W^2(r) \simeq \frac{1}{2r^2} \Big(\beta_3 + \frac{1}{\lambda} \Big)~.
\eea
From large $r$ analysis, this solution, together with (\ref{epsilon0zeta}), is also consistent with the third Einstein equation (\ref{eq:25}).

 The asymptotic analysis for $r \to \infty$ does not determine the constant $\beta_3$. 
However, from (\ref{wsigma}), which is valid for all $r$, we are led to identify 
\bea\label{p0beta3} 
\beta_3 + \frac{1}{\lambda} = 2\, \varsigma^2~, 
\eea
where on the right-hand-side we have used (\ref{torsionH2}).

We remark that, in view of (\ref{eq:28}), and that asymptotically (for $r \to \infty$)
we have $A B \simeq 1$, $B = 1 - \eta^2$ (\emph{cf.} (\ref{larger}), (\ref{epsilon0zeta})), 
the leading behaviour, for asymptotically large ($r \to \infty$), of the radial component of the magnetic field is still given by (\ref{magmon}).
On the other hand, the asymptotic behaviour for $r \to \infty$ of the axion field $b$ is 
\bea\label{baxion}
b(r) \simeq - \frac{\varsigma}{r} + \dots ~, \quad r  \to \infty~,
\eea
where we used the shift symmetry in the action (\ref{eq:13}), $b \rightarrow b + c_0$, with $c_0$ a constant, 
to impose the boundary condition $b(\infty)=0$.Thus $-\varsigma$ plays the r\^ole of an `axion' charge. In this sense the fact that $\varsigma^2 $ contributes to the ``charge''  term in the metric function $B$ (\ref{larger}), is consistent with the findings of \cite{lee}. However, our model and monopole solution are quite different from those of \cite{lee}. Moreover, in our case, asymptotically for large $r$, there is the deficit $\eta^2$ in the metric function $B$ due to the presence of the global monopole.

Secondly, for very large coupling $\lambda \to \infty$, which is of phenomenological relevance as it enforces the scalar triplet field to take on its classical vacuum expectation value, the $1/\lambda$ terms on the left-hand-side of (\ref{p0beta3})  can be ignored. 
In this case, the metric function $B$ is given by the Reissner-Nordstrom form (\ref{RN}) for \emph{both} small and large $r$.  

Finally, we remark that the $1/r$-term in $B$ in (\ref{larger}) corresponds to the contributions from the monopole mass, and has a coefficient $\beta_2$ which cannot be determined 
in our asymptotic analysis up to $O(1/r^3)$. From the expected  large $r$ asymptotic  RN form (\ref{RN}) of the metric tensor, one can identify $\beta_2 = -2\mathcal M$, where $\mathcal M$ is the monopole mass~\cite{vilenkin}.  An estimate of the monopole mass is given in the next subsection.

\subsection{An estimate of the magnetic monopole mass \label{sec:rnmass}}

To make an estimate of the monopole mass, we shall use the analytic form of the solution in the two asymptotic regimes of small and large (radial) distances from the monopole centre. The monopole mass is concentrated in the core region whose size we will estimate following arguments similar to those in ref.~\cite{vilenkin}.  The total energy (i.e. (rest) mass $\mathcal{M}$) is given by the integral over three space of the time-time component of the stress energy tensor: 
\bea\label{stresstensor}
 \mathcal{M} = \int \sqrt{-g}\, d^3x \, \Big[ \frac{2\, W^2}{B\,r^2} + \frac{(b^\prime)^2}{4\, B A}     + \eta^2\, \Big(\frac{f^2}{B  r^2} + \frac{(f^\prime)^2}{2B A} \Big) + \frac{\lambda \, \eta^4}{4 B}\, (f^2 - 1)^2 \Big]~.
\eea
From the metric (\ref{eq:19}), and the property (\ref{eq:AB}), we observe that the integration measure $\sqrt{-g}\, d^3x = r^2 {\rm sin}\theta \, dr \, d\theta \, d\phi $ in spherical polar coordinates assumes its flat space-time form. 
Taking into account the small-$r$ form of the various functions appearing in (\ref{stresstensor}), we obtain that, for $r \to 0$, the corresponding contributions to the integral are vanishing to leading order. 
However, for  $r \to \infty $, there is a linearly  divergent contribution in $r$ coming from the third term of the integrand on the right-hand-side of (\ref{stresstensor}), which is the dominant contribution to the integral. We have assumed that the interpolating functions of the various terms are non-singular in the non-asymptotic regions. Using a spatial infrared cutoff $L$,  we estimate the mass of the monopole to be
\bea\label{mass}
\mathcal{M} \sim 4\pi \, \frac{\eta^2}{1-\eta^2} \, \int_0^L dr \sim 4\pi \, \eta^2 \, L ~,
\eea
for $\eta \ll 1$ (or in terms of the dimensionful quantities (\ref{scaled}) $\eta \ll M_{\rm Pl}$, where $M_{\rm Pl}$ is the reduced Planck mass); this estimate is consistent phenomenologically (see below). Physically, and following the logic of ref.~\cite{vilenkin}, which discusses self-gravitating global monopoles in the absence of both electromagnetic fields and (Kalb-Ramond) torsion,  
we may assume that the mass of the monopole is concentrated in its core, whose size is $L$, and outside this the scalar field configuration approaches its constant vacuum expectation value, that is $f \sim 1$.

It has been estimated in \cite{vilenkin} that the core size is of order $\lambda^{-1/2}\, \eta^{-1} $ in flat space. If we replace $L$ by the core size in (\ref{mass}), then we obtain  $\mathcal{M} \sim 4\pi \lambda^{-1/2} \eta$ (the same order for the mass of the monopole given in \cite{vilenkin}).
For small $\eta \, ( \ll 1)$, gravity is expected not to change significantly the structure of the monopole at small distances. 

However, in our case we see from (\ref{core}) that the above estimate for the core size is not correct, in the sense that at such distances $f \simeq 0$ and the approximation that $f \simeq 1$ at large $r$ is not valid.
For a consistent picture,  $L$ must be such that $L \gg \sqrt{\alpha_1}= \lambda^{-1/2}\, \eta^{-1}$, so that $ f \simeq 1$.
It is sufficient to take the size of the core to be
\bea\label{coresize}
L = \xi \lambda^{-1/2}\, \eta^{-1} 
\eea
with $\xi = O(10)$ say. 

In such a case, from (\ref{mass}) we obtain the following order of magnitude estimate of the monopole mass
\bea\label{massfinal}
\mathcal{M} \sim 4\pi \, \xi \, \lambda^{-1/2} \, \eta ~, \quad \xi = \mathcal{O}(10)
\eea
with $\lambda$ and $\eta$  phenomenological parameters to be constrained by experiment. 
We will attempt to estimate the order of the parameter $\xi$ by \emph{assuming} that the main contribution to the integral (\ref{stresstensor})
comes from a thin shell of radius $R \sim L \gg 1 $, and of thickness $\Delta L = (1-\alpha )L $, $0< \alpha < 1$; we then find for $ \eta \ll 1$ and large $\lambda$ that  
\bea\label{corer1}
\mathcal{M} &\sim & \int_{\rm shell ~thickness~(1-\alpha)L}\, \sqrt{-g}\, d^3x \, \Big[ \frac{2\, W^2}{B\,r^2} + \frac{(b^\prime)^2}{4\, B A}     + \eta^2\, \Big(\frac{f^2}{B  r^2} + \frac{(f^\prime)^2}{2B A} \Big) + \frac{\lambda \, \eta^4}{4 B}\, (f^2 - 1)^2 \Big] \nonumber \\
&\simeq &  \frac{1}{\alpha}\, (1-\alpha) \, \Big( 9\pi \varsigma^2 + \frac{4\pi}{\lambda} \Big) \,  \frac{1}{L} + 4\pi \, \eta^2 \, (1 - \alpha ) \, L\Big] ~,
\eea
 which we assume here.  In arriving at the above result we have used (\ref{wsigma}), (\ref{eq:28}), and the asymptotic behaviour  for large $r$: $f^2 -1 \simeq -\frac{2}{\lambda \eta^2 r^2}$ and $AB \simeq 1$. 
 For  $\lambda \gg 1$ assumed here, which ensures that the scalar fields $\chi^a$ approach their vacuum expectation values,  
the right-hand-side of (\ref{corer1}) is practically independent of the coupling $\lambda$~\footnote{In fact, the $1/\lambda$ corrections on the right hand side of (\ref{corer1}) are absent if one defines the shell radius $L$ as the one signifying a region of space outside which one substitutes the value $f=1$ for the scalar field configuration (i.e. the fields are replaced by their vacuum expectation values). The difference in the estimate of the core size from the approach based on (\ref{corer1}) is then negligible for $\lambda \gg 1$.}.
 The core radius $L_c$   then is estimated by minimizing $\mathcal M$ as given in (\ref{corer1}) with respect to $L$, which yields 
\bea\label{coreradius}
L_c = \frac{3}{2} \, \sqrt{\frac{1}{\alpha}}\,\frac{\varsigma}{\eta}~.
\eea
 
Comparing with (\ref{coresize}), this estimate yields 
$\xi = \frac{3}{2}\sqrt{\frac{\lambda}{\alpha}}$, which can be arranged to be of O(10). This yields an estimate of the monopole mass 
\bea\label{massfinal2}
\mathcal M \sim 12\pi\, \, \sqrt{\frac{1}{\alpha}} \, (1-\alpha) \, \varsigma \, \eta~.
\eea
The important point to notice is that the mass is proportional to the Kalb-Ramond field strength (``torsion charge'') $\varsigma$ and  independent of $\lambda$ (in leading order for large $\lambda$) . Within our phenomenological effective theory, $\xi$ (equivalently $\alpha$) cannot be completely determined without a full interpolating solution.

 Before proceeding further we would like to make some comments regarding the nature of the mass in (\ref{massfinal2}) which is positive, in contrast to the discussion in \cite{negative}; thus our monopole is an ordinary particle and can be produced at a collider such as the LHC. This case has to be contrasted with the solution of \cite{negative} which, as mentioned in section \ref{sec:2}, is associated with a matching (at the core radius) of an exterior  Schwarzschild-like metric to an interior local de Sitter metric. Such a construction leads to a \emph{negative} mass for the monopole, as we have discussed, which is not of  relevance to collider physics, although there may be some cosmological interest. 

However in our case, the space-time at the origin $ r \to 0$ (\ref{eq:32})  is \emph{not} of de Sitter type  but rather of Reissner-Nordstr\"om type (\ref{RN}) 
with $B$ scaling like $1/r^2$ (\ref{eq:32}), owing to $\varsigma$ being non-zero (see (\ref{eq:28})). In our analysis we assumed that the entirety of the mass of the monopole is enclosed inside the core radius $L_c$  (\ref{coreradius}), which implies a sort of `bag' model. For $\eta \ll 1$, which we assume, the mass of the monopole is much smaller than the Planck scale. In such a case there are no Reissner-Nordstr\"om horizons, defined by the vanishing of the metric function $B$, i.e. 
\be\label{rnhor}
1 - \frac{2{\mathcal M}}{r} + \frac{2\,\varsigma^2}{ r^2} =0 ~, \Rightarrow r^{(\pm}) = {\mathcal M} \Big(1 \pm \sqrt{1 - \frac{2\,\varsigma^2}{{\mathcal M}^{2}}}\Big)~;
\ee
existence of horizons would require that ${\mathcal M} \ge \sqrt{2}\,\varsigma $ which is incompatible with (\ref{massfinal2}), since $\alpha = {\mathcal O}(1)$ and $\eta \ll M_{P}$. This means that a black hole and the corresponding horizons cannot form in our case, but the naked singularity at $r \to 0$ is still shielded inside the core radius, which essentially separates an outer region in space, where the scalar field is locked into its vacuum expectation value, from an inner region where symmetry breaking is not complete, and in fact the field $f$ vanishes at the centre $r=0$.   
On the other hand, the asymptotic form (\ref{larger}) for large but finite $r$  is of Reissner-Nordstr\"om type (\ref{RN}). As we have already mentioned, the
$1/r$-term corresponds to the contributions from the monopole mass, and has a coefficient $\beta_2$ which cannot be determined 
in our asymptotic analysis up to $O(1/r^3)$, but could be related to the monopole mass.  These are crucial features of the solution ensuring a positive mass (\ref{massfinal}) for the monopole. The $1/r^2$ deviations (\ref{larger}), (\ref{p0beta3})  from the Schwarzschild form occur due to the Kalb-Ramond axion-like field and the interactions of the Higgs field. This evasion of  Birkhoff's theorem is permitted because in the exterior of the monopole core the space-time is not that of the vacuum, being characterised by non-zero gauge and axion fields that contribute to the magnetic charge contribution of the Reissner-Nordstr\"om solution (\ref{RN}). 
\subsection{Magnetic Charge Quantization and Discrete Kalb-Ramond field strength}

An important property, of our magnetic monopole solution is the quantization of its magnetic charge. (We shall restrict ourselves to the case of very strong self interactions among the scalar fields $\lambda \to \infty$.) As we have discussed above, from the form of $W(r)$ given in (\ref{wsigma}) we deduce the following form of the (radial) magnetic field (\ref{magnetic}): 
\bea\label{magnfield}
\mathcal{\mathbf{B}} =\sqrt{2} \, \varsigma  \, \frac{\mathbf{r}}{r^3}~.
\eea
This has the same ``Coulomb-like'' form $\mathbf{B} = g \frac{\mathbf{r}}{r^3}$ of the standard Dirac monopole magnetic field~\cite{diracmono} with ``magnetic charge'' $g$ given by (\emph{cf.} (\ref{magcharge})):
\bea\label{magneticcharge}
g = \sqrt{2}\, \varsigma~.
\eea
The magnetic charge is proportional to the Kalb-Ramond  field strength.  Since the latter can be positive or negative, the charge can be positive or negative, which implies the existence of both monopoles and antimonopoles. 

Topological quantization of the magnetic charge ($\varsigma$ in our case)   found in the standard 't Hooft-Polyakov monopole solution \cite{hpmono}, does not follow from the $H$-dependent 
modification of the electromagnetic tensor (\ref{hpfmn}). This is because  the latter is four-dimensional in the internal space: there is a tetrad of scalar fields $\phi^a$, $a=0,1,2,3$ (\ref{hpfmn}) which does not provide a mapping of an internal SO(3) sphere onto a spatial $S^2$ sphere. 
 
The quantization of $\varsigma$ in our case can \emph{only} come from the standard Dirac argument~\cite{diracmono}, which  considers the gauge transformations of the quantum relativistic wavefunction $\psi$ of an electron field (with electric charge $e$) 
in the presence of the Dirac-string singular vector potential $\mathbf{A}(\mathbf{r})$ for the monopole magnetic field $\mathbf{B}$ (with  $\mathbf{B} = \mathbf{\nabla} \times \mathbf{A} $). Explicitly, requiring the single-valuedness of the wave function under the appropriate (singular) gauge transformations, yields the Dirac quantization rule
\bea\label{rule}
 |g\, e| = \frac{n}{2}~, \quad n \in {\mathbf Z^+} \, {\cup} \, \{0\}.
 \eea
 From (\ref{magneticcharge}), we obtain the discretization of the Kalb-Ramond  field strength
 \bea\label{discretetorsion}
 \varsigma \, e = \frac{n}{2\, \sqrt{2}}~, \quad n \in {\mathbf Z^+}  \, { \cup} \, \{0\}.~.
 \eea
The discreteness of the Kalb-Ramond field strength in the presence of a magnetic monopole is a novel feature of our solution~\footnote{We should point out that discrete Kalb-Ramond field strengths may also arise in certain bosonic $\sigma$-models where the target space time is a group manifold~\cite{aben}, and the corresponding field strength is proportional to the level of the associated Kac-Moody algebra. Such models perhaps might provide a framework for embedding our solution in an ultraviolet-complete model of string theory.}. 
 
We should also note that the antisymmetric-tensor monopole solutions of \cite{nepo}, differ from ours in several ways. Firstly, those monopoles are considered in $D$-dimensional Minkowski space time ($D=n + 3$, $n \ge 2$ a positive integer). Secondly, as a result of appropriately fixing the antisymmetric tensor gauge symmetry, and taking the $B$-field to be time-independent, the static field strength $H_{\mu_1\dots \mu_{n+1}} = \partial_{[\mu_1}B_{\mu_2\dots \mu_{n+1}]}$  can be considered as a form in a $D-1$ Euclidean space, which can be patched appropriately in the presence of an antisymmetric-tensor monopole singularity at the spatial origin, leading to the quantization of the $H$-charge.
By contrast, in our solution,  
the non-zero components of the Kalb-Ramond field strength read, for the asymptotic regions $r \to 0$ or $r \to \infty$:
 \bea\label{htorsion} 
  H_{0 \theta \, \phi} = \epsilon_{0\, \theta\, \phi\, r} \,\partial^r b \, \sim  \, \varsigma \, {\rm sin}\, \theta ~, 
  \eea
 corresponding to a \emph{time-dependent} $B$-field in those regions of the form:
 \begin{equation}
B_{\mu\nu}=\left(\begin{array}{cccc}
  0 & 0 & 0 & 0\\
   0 & 0 & 0 & 0\\
   0 & 0 & 0 & \varsigma \, t\, \sin\theta\,  \\
    0 & 0 & - \varsigma \, t\, \sin\theta\, & 0
      \end{array}\right)\label{eq:bfield}
   \end{equation} 
 with $t$ the time. The field strength (\ref{htorsion}) is regular as $r \to 0$, in contrast to the solutions of ref.~\cite{nepo}, where the $H$-charge quantization was a consequence of monopoles in the antisymmetric tensor field, with singular behaviour of the $H$-field strength as $r \to 0$.  
  
\section{Discussion: Constraining the model by Experiment  \label{sec:5}}

We have outlined how Kalb-Ramond axion fields could
generate a magnetic monopole. There is spontaneous symmetry breaking
of a global internal symmetry which is essential for producing a self-gravitating global 
monopole with a deficit angle. In the limit $\lambda \to \infty$
the scalar fields have no propagating degrees of freedom and are confined to classical values that interpolate
between the two expectation values that minimize the potential, that is between zero and $\eta$. Without  non-zero  Kalb-Ramond field strengths (``torsion'') ,
we showed that no magnetic monopole charge is induced by the global monopole (in the limit of large  $\lambda$ but there are ${\mathcal O}(1/\lambda)$ contributions of course). Because
non-singular abelian gauge fields are incompatible with magnetic
monopoles, we have worked directly with the electromagnetic field tensor and
not required the abelian Bianchi identity.
Nevertheless, formally, we can use a singular vector potential to represent the magnetic field, which has the Coulomb-like form of the 
standard Dirac monopole.  Following the argument of Dirac the latter is  quantised in terms of the fundamental electric charge. 

The existence of the magnetic charge (\ref{magneticcharge}) implies high ionization, while the smallness of the  monopole mass (\ref{massfinal}) (as compared to the Planck scale) makes our magnetic monopole model falsifiable in the current round of the LHC~\cite{monodata}. The scalars $\chi^a$ (\ref{eq:22})
in the model do \emph{not} represent the Standard Model Higgs, but elementary defects (or composites of heavy fermions) associated with a spontaneous breaking of  O(3) symmetry~\cite{vilenkin}; the phenomenological parameters $\lambda$ and the vacuum expectation value $\eta$, as well as the core size parameter $\xi$ can be  constrained by experiments, if one accepts the loose definition of the core in (\ref{coresize}) which leads to (\ref{massfinal}); the core dimension is some large distance (compared to the Planck length) such that, 
for distances larger than this, the solution for the scalar field configuration is $f \simeq 1$ (up to terms of O($\xi^{-1}\lambda^{-1/2}\eta^{-1}$)).  

It is important to note that large $\lambda$ self-interaction couplings 
affect significantly the probability for producing monopole-antimonopole pairs~\cite{vilenkin}, 
\bea\label{prob}
\mathcal{P} \propto e^{-({\rm const}) \mathcal{M}^2/F } \propto e^{-{\rm const}^\prime /\lambda}, 
\eea
where $F \sim \eta^2$ is the attractive force between a monopole and an antimonopole due to the linear divergent energy. In fact this can be understood by estimating the form factor in the cross-section for such processes. 
Indeed, from (\ref{coresize}) and (\ref{massfinal}) one may estimate the ratio of the core size $R_{\rm core} = L$ to the Compton wavelength $\lambda_{\rm Compt} = 1/\mathcal{M}$ of the monopole as
\bea\label{ratio}
\frac{R_{\rm core}}{\lambda_{\rm Compt}} \sim 4\pi \, \xi^2 \, \lambda^{-1} = \frac{\mathcal{M}^2}{4\pi\, \eta^2}\quad.
\eea
For large monopoles $R_{\rm core} > \lambda_{\rm Compt}$, and therefore weak self-interaction couplings $\lambda \ll 4\pi \xi^2 = \mathcal{O}(10^3)$ for $\xi =\mathcal{O}(10)$, a semi-classical situation is reached where the form factor has an exponential suppression~\cite{nussinov} $e^{-4 R_{\rm core}/\lambda_{\rm Compt}}$. This is in agreement with the estimate in \cite{vilenkin} for the production probability $\mathcal{P}$ of monopole-antimonopole pairs (\ref{prob}). 
Consqeuently, for small $\lambda < 1$  the production probability $\mathcal{P}$ appears to be negligibly small. 
For large $\lambda \gg 1$, however, which is the case of interest here, 
$\mathcal{P}$ is expected to be large, and thus of relevance to collider (including LHC) phenomenology (although  strong coupling can complicate analytical calculations). On the other hand, once a monopole/antimonopole pair is produced, energy losses to Goldstone fields associated with the breaking of the O(3) symmetry, at a rate of order $\eta^2$~\cite{vilenkin} are probably expected, which must be taken into account when considering the relevant phenomenology. We  also have in our model photons and Kalb-Ramond fields which couple to the monopoles gravitationally and complicate the situation.

In the present article we note that from the current bounds on the (scalar) monopole mass at the LHC~\cite{monodata}, we obtain from
(\ref{massfinal2}) for the lowest magnetic charge: 
\bea\label{bounds1}
6\pi \sqrt{\frac{1}{2\,\alpha}} \, (1-\alpha)  \, \eta  \ge 420 \, ~{\rm GeV} \quad ({\rm for~spin} ~0, \, \varsigma =\frac{1}{2\sqrt{2}}, \, \lambda \gg 1, \eta \ll M_P, \alpha ={\mathcal O}(1)),   
\eea
with higher bounds for higher magnetic charges. However we should exercise great caution in applying the above limits to our model. These bounds have been derived based on perturbative Drell-Yan processes~\cite{monodata}, from the decay of virtual photons into monopole-antimonopole pairs. In the presence of strong magnetic charges, such bounds are not strictly valid. Moreover, in our model, the production mechanism of the global monopoles from standard model particle collisions needs to be carefully evaluated. In an effective field theory framework~\cite{fermi2,fermi}, the derivative of the Kalb-Ramond axion field $\partial_\mu b$ couples to the axial fermion current $\psi_i \gamma^\mu \gamma^5 \psi_i$, where $i$ runs over fermion species of the Standard Model (SM),  but the scalars couple only gravitationally to the SM fields (unless they are composites of heavy fermions which may couple to SM fields through loops). Hence the actual production mechanism of the global monopole solution at colliders, such as the LHC, and their detection, needs to be examined carefully. We stress once again that the monopole production is certainly expected to be strongly suppressed unless  $\lambda$ is  large, which is our case.

Before closing we would like to make two important additional remarks. Given that our monopole mass is of order TeV, and thus much smaller than the Planck mass $M_P$, gravitational collapse to a black hole is not expected. Indeed, it is known in general~\cite{gibbons} that to form an Abelian black hole of Arnowitt-Deser-Misner (ADM) mass $M_{\rm ~BH}$ and magnetic charge $g$, one needs to satisgfy the condition 
$$ M_{\rm BH} \ge \frac{g}{\sqrt{4\pi G_N}}~. $$
Since in our monopole case ${\mathcal M} \sim \eta g$ (\emph{cf.} (\ref{massfinal2})), in order to have a collapse one would need $ \eta \ge  \frac{M_P}{2\sqrt{\pi}}$, which is not the case, as $\eta$ is assumed in our model to be of order TeV. 

Moreover, we did not comment here on the stability of our monopole configurations. Although topologically non-trivial, indicating stability on generic grounds, there is  nevertheless an ongoing debate~\cite{debate} on the stability of global monopoles of \cite{vilenkin}, which may be subject to a sort of angular collapse. As stressed by Achucarro and Urrestilla in \cite{debate}, in the case of a global monopole, the energy barrier between the monopole and the vacuum is finite, despite the existence of a conserved topological charge and this feature is independent of the details of the scalar potential. But the issue of decay of such configurations, e.g. due to thermal fluctuation instabilities, remained inconclusive. 
Several extensions of the original model have been suggested in the literature~\cite{debate}. The extension may involve more scalars, perhaps, for instance, gauged ones  in addition to the global $\chi^a$ fields~\cite{achu}. We shall not discuss  such issues further here.

In our case, the presence of the Kalb-Ramond antisymmetric tensor and gauge  fields, which leads to real magnetic monopoles, makes the model different from others in the literature. The stability of the monopoles  deserves further studies. Even if there are instabilities, the collider production of unstable monopoles would lead to novel experimental signatures from the decay of such objects to Goldstone bosons and other particles. There are also, of course, many other interesting questions which we have mentioned and hope to address in the future.

\acknowledgments 

We thank Malcolm Fairbairn, Jim Pinfold and fellow members of the MoEDAL Collaboration  for discussions and their interest in this work. This work is supported in part by the U.K.~Science and Technology Facilities
Council (STFC) via the Grant ST/L000326/1.


\begin{thebibliography}{1}

\bibitem{monodata} G.~Aad {\it et al.} [ATLAS Collaboration],
  Phys.\ Rev.\ D {\bf 93}, no. 5, 052009 (2016)
  doi:10.1103/PhysRevD.93.052009
  [arXiv:1509.08059 [hep-ex]];
B.~Acharya {\it et al.} [MoEDAL Collaboration],
  arXiv:1604.06645 [hep-ex].



\bibitem{Hammond:2002rm} D.~J.~Gross and J.~H.~Sloan,
  Nucl.\ Phys.\ B {\bf 291}, 41 (1987).
  doi:10.1016/0550-3213(87)90465-2
R.~R.~Metsaev and A.~A.~Tseytlin,
  Nucl.\ Phys.\ B {\bf 293}, 385 (1987).
  doi:10.1016/0550-3213(87)90077-0.
See also:
  R.~T.~Hammond,
  Rept.\ Prog.\ Phys.\  {\bf 65}, 599 (2002).
  doi:10.1088/0034-4885/65/5/201, and references therein. 
  
  
  
\bibitem{string}  J.~Polchinski,
  \emph{String theory. Vols. 1 and 2} (Cambridge University Press 2007)
   ISBN: 9780511252273 (eBook), 9780521672276 (Print), 9780521633031 (Print);
   ISBN: 9780511252280 (eBook), 9780521633048 (Print), 9780521672283 (Print).




\bibitem{diracmono} P.~A.~M.~Dirac,
  Phys.\ Rev.\  {\bf 74}, 817 (1948).
  doi:10.1103/PhysRev.74.817;
  Int.\ J.\ Theor.\ Phys.\  {\bf 17}, 235 (1978).
  doi:10.1007/BF00672870

\bibitem{hpmono} G.~'t Hooft,
  Nucl.\ Phys.\ B {\bf 79}, 276 (1974).
  doi:10.1016/0550-3213(74)90486-6;
  A.~M.~Polyakov,
  JETP Lett.\  {\bf 20}, 194 (1974)
  [Pisma Zh.\ Eksp.\ Teor.\ Fiz.\  {\bf 20}, 430 (1974)].


\bibitem{gg} H.~Georgi and S.~L.~Glashow,
  Phys.\ Rev.\ Lett.\  {\bf 32}, 438 (1974).
  doi:10.1103/PhysRevLett.32.438


\bibitem{largedim} N.~Arkani-Hamed, S.~Dimopoulos and G.~R.~Dvali,
  Phys.\ Lett.\ B {\bf 429}, 263 (1998)
  doi:10.1016/S0370-2693(98)00466-3
  [hep-ph/9803315];
I.~Antoniadis, N.~Arkani-Hamed, S.~Dimopoulos and G.~R.~Dvali,
  Phys.\ Lett.\ B {\bf 436}, 257 (1998)
  doi:10.1016/S0370-2693(98)00860-0
  [hep-ph/9804398];
L.~Randall and R.~Sundrum,
  Phys.\ Rev.\ Lett.\  {\bf 83}, 3370 (1999)
  doi:10.1103/PhysRevLett.83.3370
  [hep-ph/9905221].


  
\bibitem{vilenkin} M.~Barriola and A.~Vilenkin,
  Phys.\ Rev.\ Lett.\  {\bf 63}, 341 (1989).
  doi:10.1103/PhysRevLett.63.341

\bibitem{fermi2} M.~de Cesare, N.~E.~Mavromatos and S.~Sarkar,
  Eur.\ Phys.\ J.\ C {\bf 75}, no. 10, 514 (2015)
  doi:10.1140/epjc/s10052-015-3731-z
  [arXiv:1412.7077 [hep-ph]].

  
\bibitem{torsionglobal} F.~Rahaman, S.~Sur and K.~Gayen,
  Phys.\ Scripta {\bf 69}, 78 (2004).
  doi:10.1238/Physica.Regular.069a00077.



\bibitem{papavassiliou} 
  P.~O.~Mazur and J.~Papavassiliou,
  Phys.\ Rev.\ D {\bf 44}, 1317 (1991).
  doi:10.1103/PhysRevD.44.1317;
  H.~Ren,
  Phys.\ Lett.\ B {\bf 325}, 149 (1994)
  doi:10.1016/0370-2693(94)90085-X
  [hep-th/9312074];
E.~R.~Bezerra de Mello and C.~Furtado,
  Phys.\ Rev.\ D {\bf 56}, 1345 (1997).
  doi:10.1103/PhysRevD.56.1345
A.~A.~Roderigues Sobreira and E.~R.~Bezerra de Mello,
  Grav.\ Cosmol.\  {\bf 5}, 177 (1999)
  [hep-th/9809212];


\bibitem{debate} A.~S.~Goldhaber,
  Phys.\ Rev.\ Lett.\  {\bf 63}, 2158 (1989).
  doi:10.1103/PhysRevLett.63.2158;
In the original suggestion of Goldhaber that global monopoles are not stable against ``angular'' collapse, there is an ongoing debate on this issue; for a partial list of references  see:
S.~H.~Rhie and D.~P.~Bennett,
  Phys.\ Rev.\ Lett.\  {\bf 67}, 1173 (1991).
  doi:10.1103/PhysRevLett.67.1173;
L.~Perivolaropoulos,
  Nucl.\ Phys.\ B {\bf 375}, 665 (1992).
  doi:10.1016/0550-3213(92)90115-R;
G.~W.~Gibbons, M.~E.~Ortiz, F.~Ruiz Ruiz and T.~M.~Samols,
  Nucl.\ Phys.\ B {\bf 385}, 127 (1992)
  doi:10.1016/0550-3213(92)90097-U
  [hep-th/9203023];
M.~Hindmarsh,
  Nucl.\ Phys.\ B {\bf 392}, 461 (1993)
  doi:10.1016/0550-3213(93)90681-E
  [hep-ph/9206229];
G.~Arreaga, I.~Cho and J.~Guven,
  Phys.\ Rev.\ D {\bf 62}, 043520 (2000)
  doi:10.1103/PhysRevD.62.043520
  [gr-qc/0001078];
A.~Achucarro and J.~Urrestilla,
  Phys.\ Rev.\ Lett.\  {\bf 85}, 3091 (2000)
  doi:10.1103/PhysRevLett.85.3091
  [hep-ph/0003145];
  R.~Gregory and C.~Santos,
  Class.\ Quant.\ Grav.\  {\bf 20}, 21 (2003)
  doi:10.1088/0264-9381/20/1/302
  [hep-th/0208037];
  E.~R.~Bezerra de Mello,
  Phys.\ Rev.\ D {\bf 68}, 088702 (2003)
  doi:10.1103/PhysRevD.68.088702
  [hep-th/0304029];
  S.~B.~Gudnason and J.~Evslin,
  Phys.\ Rev.\ D {\bf 92}, no. 4, 045044 (2015)
  doi:10.1103/PhysRevD.92.045044
  [arXiv:1507.03400 [hep-th]].
  
  \bibitem{negative} D.~Harari and C.~Lousto,
  Phys.\ Rev.\ D {\bf 42}, 2626 (1990).
  doi:10.1103/PhysRevD.42.2626

\bibitem{bronnikov} K.~A.~Bronnikov, B.~E.~Meierovich and E.~R.~Podolyak,
  J.\ Exp.\ Theor.\ Phys.\  {\bf 95}, 392 (2002)
  [Zh.\ Eksp.\ Teor.\ Fiz.\  {\bf 122}, 459 (2002)]
  doi:10.1134/1.1513811
  [gr-qc/0212091].



\bibitem{aben} I.~Antoniadis, C.~Bachas, J.~R.~Ellis and D.~V.~Nanopoulos,
  Phys.\ Lett.\ B {\bf 211}, 393 (1988).
  doi:10.1016/0370-2693(88)91882-5
  
  \bibitem{halpern} M.~B.~Halpern,
  Phys.\ Rev.\ D {\bf 19}, 517 (1979).
  doi:10.1103/PhysRevD.19.517


\bibitem{cherneff} P.~Das, P.~Jain and S.~Mukherji,
  Int.\ J.\ Mod.\ Phys.\ A {\bf 16}, 4011 (2001)
  doi:10.1142/S0217751X01004840
  [hep-ph/0011279].
S.~Kar, P.~Majumdar, S.~SenGupta and A.~Sinha,
  Eur.\ Phys.\ J.\ C {\bf 23}, 357 (2002)
  doi:10.1007/s100520100872
  [gr-qc/0006097].
D.~Maity, S.~SenGupta and S.~Sur,
  Phys.\ Rev.\ D {\bf 72}, 066012 (2005)
  doi:10.1103/PhysRevD.72.066012
  [hep-th/0507210].
J.~Alexandre, N.~E.~Mavromatos and D.~Tanner,
  Phys.\ Rev.\ D {\bf 78}, 066001 (2008)
  doi:10.1103/PhysRevD.78.066001
  [arXiv:0804.2353 [hep-th]];
  New J.\ Phys.\  {\bf 10}, 033033 (2008)
  doi:10.1088/1367-2630/10/3/033033
  [arXiv:0708.1154 [hep-th]].

\bibitem{fermi} M.~J.~Duncan, N.~Kaloper and K.~A.~Olive,
  Nucl.\ Phys.\ B {\bf 387}, 215 (1992).
  doi:10.1016/0550-3213(92)90052-D;



  
  \bibitem{RN} Sean M. Caroll,  \emph{Spacetime and Geometry} (Addison Wesley, 2004);
C. Misner, K. Thorne, and J. Wheeler, \emph{Gravitation } Freeman, 1973). 
  
  
  \bibitem{lee} K.~M.~Lee and E.~J.~Weinberg,
  Phys.\ Rev.\ D {\bf 44}, 3159 (1991).
  doi:10.1103/PhysRevD.44.3159
  
  
  \bibitem{sen} A.~Sen,
  Phys.\ Rev.\ Lett.\  {\bf 69}, 1006 (1992)
  doi:10.1103/PhysRevLett.69.1006
  [hep-th/9204046].

  \bibitem{axionbh} See for instance: S.~Sur, S.~Das and S.~SenGupta,
  JHEP {\bf 0510}, 064 (2005)
  doi:10.1088/1126-6708/2005/10/064
  [hep-th/0508150];
  T.~Ghosh and S.~SenGupta,
  Phys.\ Lett.\ B {\bf 678}, 112 (2009)
  doi:10.1016/j.physletb.2009.05.063
  [arXiv:0906.0686 [hep-th]];
  Phys.\ Lett.\ B {\bf 696}, 167 (2011)
  doi:10.1016/j.physletb.2010.12.016
  [arXiv:1101.1736 [gr-qc]] and references therein.
  
  
  
\bibitem{nepo} R.~Savit,
  Phys.\ Rev.\ Lett.\  {\bf 39}, 55 (1977).
  doi:10.1103/PhysRevLett.39.55;
P.~Orland,
  Nucl.\ Phys.\ B {\bf 205}, 107 (1982).
  doi:10.1016/0550-3213(82)90468-0;
R.~I.~Nepomechie,
  Phys.\ Rev.\ D {\bf 31}, 1921 (1985).
  doi:10.1103/PhysRevD.31.1921

\bibitem{nussinov} A.~K.~Drukier and S.~Nussinov,
  Phys.\ Rev.\ Lett.\  {\bf 49}, 102 (1982).
  doi:10.1103/PhysRevLett.49.102




\bibitem{gibbons} G.~W.~Gibbons,
  Lect.\ Notes Phys.\  {\bf 383}, 110 (1991)
  doi:10.1007/3-540-54293-024
  [arXiv:1109.3538 [gr-qc]], and references therein; 
see also: K.~M.~Lee, V.~P.~Nair and E.~J.~Weinberg,
  Phys.\ Rev.\ D {\bf 45}, 2751 (1992)
  doi:10.1103/PhysRevD.45.2751
  [hep-th/9112008].


\bibitem{achu} A.~Achucarro, B.~Hartmann and J.~Urrestilla,
  JHEP {\bf 0507}, 006 (2005)
  doi:10.1088/1126-6708/2005/07/006
  [hep-th/0504192];

\end{thebibliography}
\end{document}